\documentclass[sn-standardnature, noinfo]{sn-jnl}

\jyear{2021}%

\theoremstyle{thmstyleone}%
\usepackage[numbers]{natbib}
\usepackage[hang]{footmisc}

\makeatletter
\def\blfootnote{\xdef\@thefnmark{}\@footnotetext} 
\makeatother

\theoremstyle{thmstyletwo}%

\theoremstyle{thmstylethree}%

\begin{document}

\title[Personalization of Web Search During the 2020 US Elections]{Personalization of Web Search During the 2020 US Elections}


\author[1]{\fnm{Ulrich} \sur{Matter}}\email{ulrich.matter@unisg.ch}

\author[1]{\fnm{Roland} \sur{Hodler}}\email{roland.hodler@unisg.ch}

\author[2]{\fnm{Johannes} \sur{Ladwig}}\email{johannes@me.com}

\affil[1]{\orgdiv{Department of Economics}, \orgname{University of St.Gallen}, \orgaddress{\street{Bodanstrasse 8}, \city{St.Gallen}, \postcode{9000}, \state{SG}, \country{Switzerland}}}

\affil[2]{\orgdiv{Department of Economics}, \orgname{London School of Economics and Political Science}, \orgaddress{\street{WC2A 2AE}, \city{London}, \country{United Kingdom}}}



\abstract{Search engines play a central role in routing political information to citizens. The algorithmic personalization of search results by large search engines like Google implies that different users may be offered systematically different information. However, measuring the causal effect of user characteristics and behavior on search results in a politically relevant context is challenging. We set up a population of 150 synthetic internet users (``bots'') who are randomly located across 25 US cities and are active for several months during the 2020 US Elections and their aftermath. These users differ in their browsing preferences and political ideology, and they build up realistic browsing and search histories. We run daily experiments in which all users enter the same election-related queries. Search results to these queries differ substantially across users. Google prioritizes previously visited websites and local news sites. Yet, it does not generally prioritize websites featuring the user's ideology. 
}

\keywords{Search engines, algorithmic personalization, political news, informational segregation, US elections}


\renewcommand*{\thefootnote}{\fnsymbol{footnote}}
\renewcommand{\hangfootparindent}{0em}
\renewcommand{\footnotelayout}{\hspace{0em}}

\footnotetext[0]{We thank Yarden Katz and Alois Stutzer for valuable feedback; Dominik Burkolter and Nermin Elkasovic for valuable contributions and support in software development; and Noel Ackermann and Beatrice Blini for excellent research assistance. Ulrich Matter gratefully acknowledges financial support from the Swiss National Science Foundation grant \#190429.}

\maketitle
\renewcommand\thefootnote{\arabic{footnote}}

\section{Introduction}\label{sec:intro}

Many large web platforms, in particular search engines, personalize the information they provide to users based on the users' characteristics and preferences \cite{hannak_2013,Hannak_et_al_2017,Robertson_et_al_2018,Krafft_2019,le_etal2019}. Personalized search results on consumption goods and entertainment choices may well be in the users' best interest. Yet, personalized search results on political news and events may be less desirable. There are concerns that users might unintentionally end up consuming only political information conforming to their point of view \cite{Pariser_2011}, while democracies tend to work better if all citizens are exposed to political information from diverse viewpoints \cite{Sunstein_2001}. However, so far, little is known about the causal effect of the search engine's algorithmic personalization on systematic differences in political search results. Filling this gap is a prerequisite for understanding whether personalization of web search could eventually lead to segregation in political information.


To fill this gap, we study the search results provided by Google Search in response to election-related queries during the 2020 US Elections and their aftermath. For this purpose, we set up a population of 150 synthetic internet users (``bots'') with different partisan preferences who are randomly located across 25 US cities and active from October 22, 2020, to February 8, 2021. Our study design rests on two cornerstones: First, our users build up realistic, but potentially partisan browsing and search histories. Second, to test for personalization of search results, we run daily experiments in which all users enter the same election-related search terms. 

The importance of location or browsing histories for the personalization of results from search engines and news aggregators has been studied before. Important previous contributions rely on short-run experiments in a controlled setting \cite{hannak_2013,Hannak_et_al_2017,le_etal2019} 
or search results received by self-selected internet users whose browsing and search histories are unknown to the researchers \cite{Robertson_et_al_2018,Krafft_2019}. 
Our approach combines the advantages of controlled experiments with the advantage of having users who display a human-like browsing and search behavior and who are entirely transparent as well as traceable over several months. 
Our study also differs from previous research by focusing on a politically turbulent and contested period, in which political news was abundant. The intense coverage of election-related events by websites across the political spectrum combined with the large share of the population searching for this information on Google Search are one potential reason for why we find stronger evidence for personalization than some recent studies \cite{Robertson_et_al_2018,Krafft_2019} (see \emph{Supplementary Material}, Sections~A and B, for details on how the context of our study helps to address common challenges in measuring the effects of personalization).  

Our study design rests on 150 user profiles that emulate different hardware/software combinations (unique device fingerprints), which allow tracking via cookies and fingerprint detection even though the users are not logged into any account. 
We randomly split these profiles into three equally sized groups of synthetic users and assign a political ideology to each group: either Democrat/liberal, Republican/conservative, or non-partisan. We use residential proxy servers to randomly assign two Democrat, two Republican and two non-partisan users to each of 25 US cities with different partisan compositions (see Methods).

Users are active 1--3 times per day between 9am and 4pm local time. Their browsing and search behavior includes different common and partisan activities (see Methods).
Each user directly visits a small random selection of popular US websites on a daily basis. Moreover, each Democrat (Republican) user gets assigned ten out of 100 liberal (conservative) websites and visits 3--5 of these websites per day. This assignment is consistent with partisans' preferences for partisan information sources \cite{iyengar_hahn2009, Flaxman_et_al_2016, peterson2018echo, peterson_goel_iyengar_2021} and implies that the personal sets of partisan websites do not only differ between Democrat and Republican users but also within Democrat users and within Republican users. As such, each non-partisan user has a fixed set of non-partisan websites (and the corresponding domains, i.e., the websites' addresses), while each partisan user has fixed sets of both non-partisan and partisan websites (and the corresponding domains). Below, we refer to these websites (domains) as a user's ``favorite'' websites and state that the user is ``familiar'' with a website if it is in their set of favorite websites.
In addition to directly visiting favorite websites, we configure the synthetic users to also use Google Search in order to find and consume both non-partisan and partisan content. Each user issues 1--2 non-partisan Google searches per day. Moreover, each Democrat (Republican) user gets assigned 60 common liberal (conservative) search terms, and launches 3--9 partisan searches per day -- consistent with partisan information seeking \cite{peterson_shanto2021}.
Users always select the first entry of the organic search results on the search engine result page presented to them.

\begin{figure}[!htb]
\centering
\includegraphics[width=.99\linewidth]{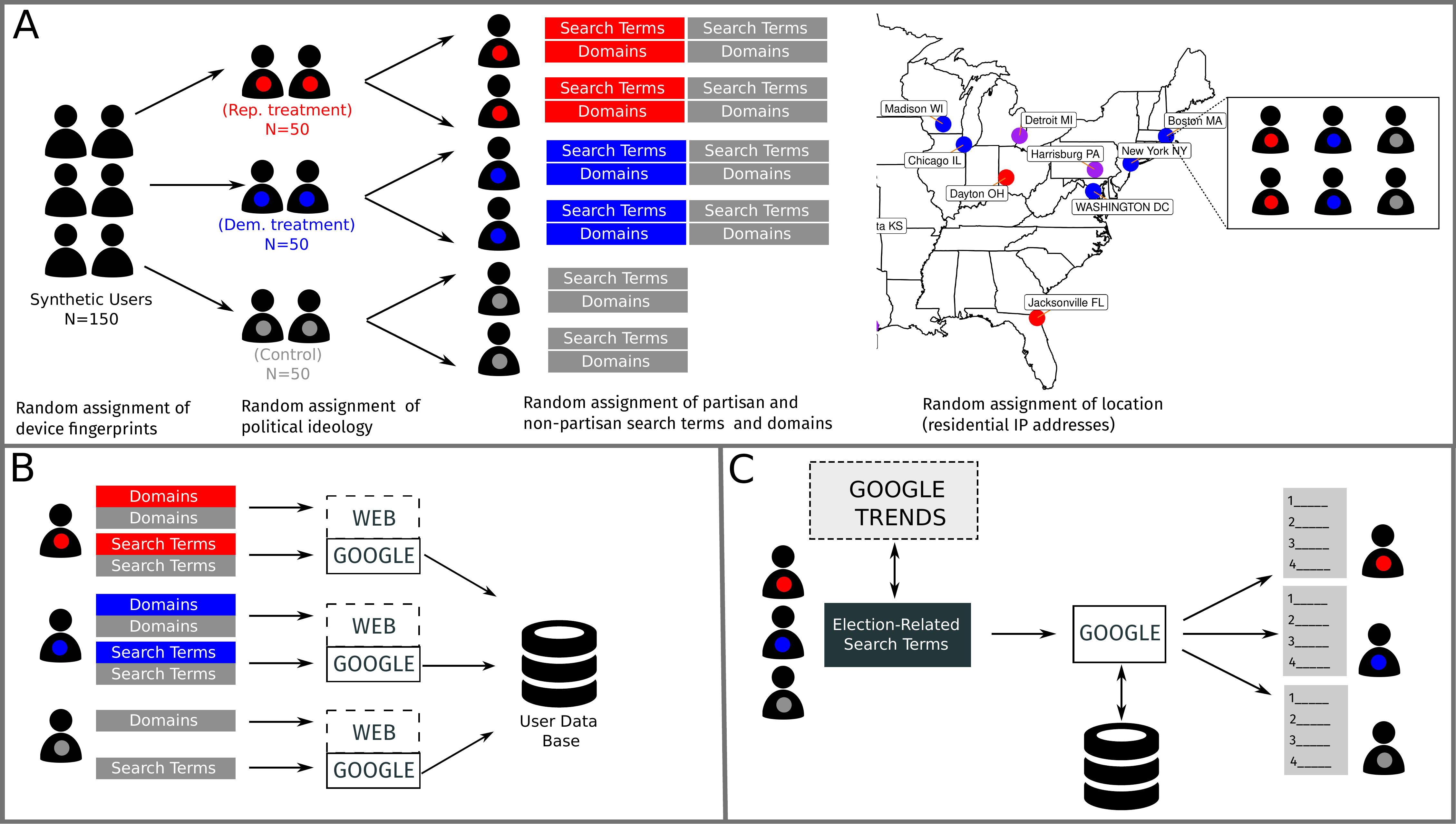}
\caption{Overview of study design. Panel A summarizes the initial setup of the synthetic user population, including (from left to right) the random assignment of device fingerprints, political ideology, search terms and domains, and location. Panel B illustrates the users' partisan and common daily browsing and search behavior, which reveals preferences and personal characteristics to Google Search. Google Search is expected to gain a detailed user data base through the user's activities on Google's websites as well as through third party tracking in the web. Panel C illustrates the identical election-related queries (sourced from Google Trends) that allow us to capture the effects of user characteristics and behavior on the user's election-related search results.}
\label{fig:design}
\end{figure}

We run daily experiments in which all users use the same election-related search terms. The pool of election-related search terms is updated throughout our study from topic pages provided by Google Trends. On these pages, Google Trends maintained lists of the most highly trending search terms related to the elections and their aftermath. 
For example, from mid-October to mid-November 2020, Google Trends had a topic page on the elections; later it had topic pages on the Capitol riots, Trump's second impeachment trial, and Joe Biden's cabinet and his first actions in office. The search terms used in our daily experiments are thus representative of what people in the US frequently googled during and after the elections, e.g., ``Donald Trump'' and ``Polling station'' right before election day, ``illegal ballots'' and ``electoral vote'' after the election, as well as ``national guard in capitol'' and ``capitol police officer dies'' after the Capitol riots. In these daily experiments, the users also always select the first entry of the organic search results. Figure~\ref{fig:design} illustrates our study design and Figure S1 in the \emph{Supplementary Material} the timeline and the geographical distribution of our users.

Our analyses build primarily on the organic search results on the first result page in response to the election-related queries. 
In additional analyses, we also look at the ``top stories'' section on the first result pages (see \emph{Supplementary Material}, Figure~S6).

\section{Results}\label{sec:results}

\subsection{Finding I: Search results differ substantially across users}

We assess the similarity of the search results across users who enter the same election-related search term on the same day. For each such search and user pair, we compute the Jaccard Index \cite{jaccard_1901} and the extrapolated rank-biased overlap (RBO) \cite{webber_etal2010} based on the search results' main domains (e.g., nytimes.com rather than https://international.nytimes.com/...). The Jaccard Index ignores the order of the items, while we take it into account with the extrapolated RBO and rank weights consistent with empirical click-through-rates (CRT; see Materials and Methods). 

\begin{figure}[bh]
\centering
\includegraphics[width=0.9\linewidth]{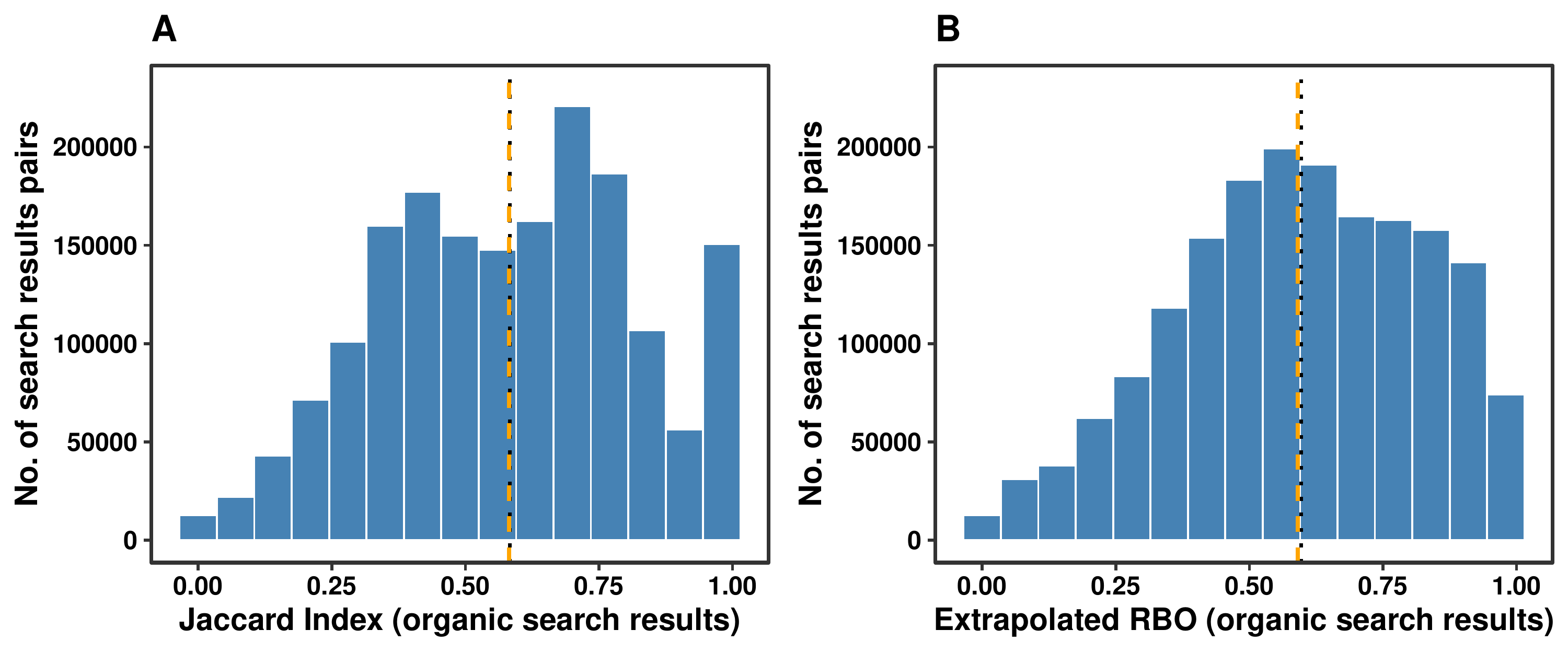}
\caption{Similarity of search results across user pairs. Histograms showing the distributions of search results similarity for each synthetic user-pair resulting from the same election-related search on the same day, measured with the Jaccard Index in panel A and the extrapolated RBO in panel B. Dotted black (dashed orange) vertical lines indicate the median (mean) of the corresponding distribution. 
} 
\label{fig:histograms}
\end{figure}

Figure~\ref{fig:histograms} reports the distributions of these two similarity measures in panels A and B. The mean of the Jaccard Index is 0.58, implying that an average user pair shares roughly 6 out of 8 results. The extrapolated RBO further reveals a remarkably low share of identical result pages (i.e., pages with the same organic search results in the same order). These findings are very similar for the top stories section on the first search results page and are robust to more or less narrow time windows between the two users' election-related queries (see \emph{Supplementary Material}, Figures S7 and S8). 



\subsection*{Finding II: Search results prioritize previously visited websites} 

\begin{figure}[tbh]
\centering
\includegraphics[width=11.4cm]{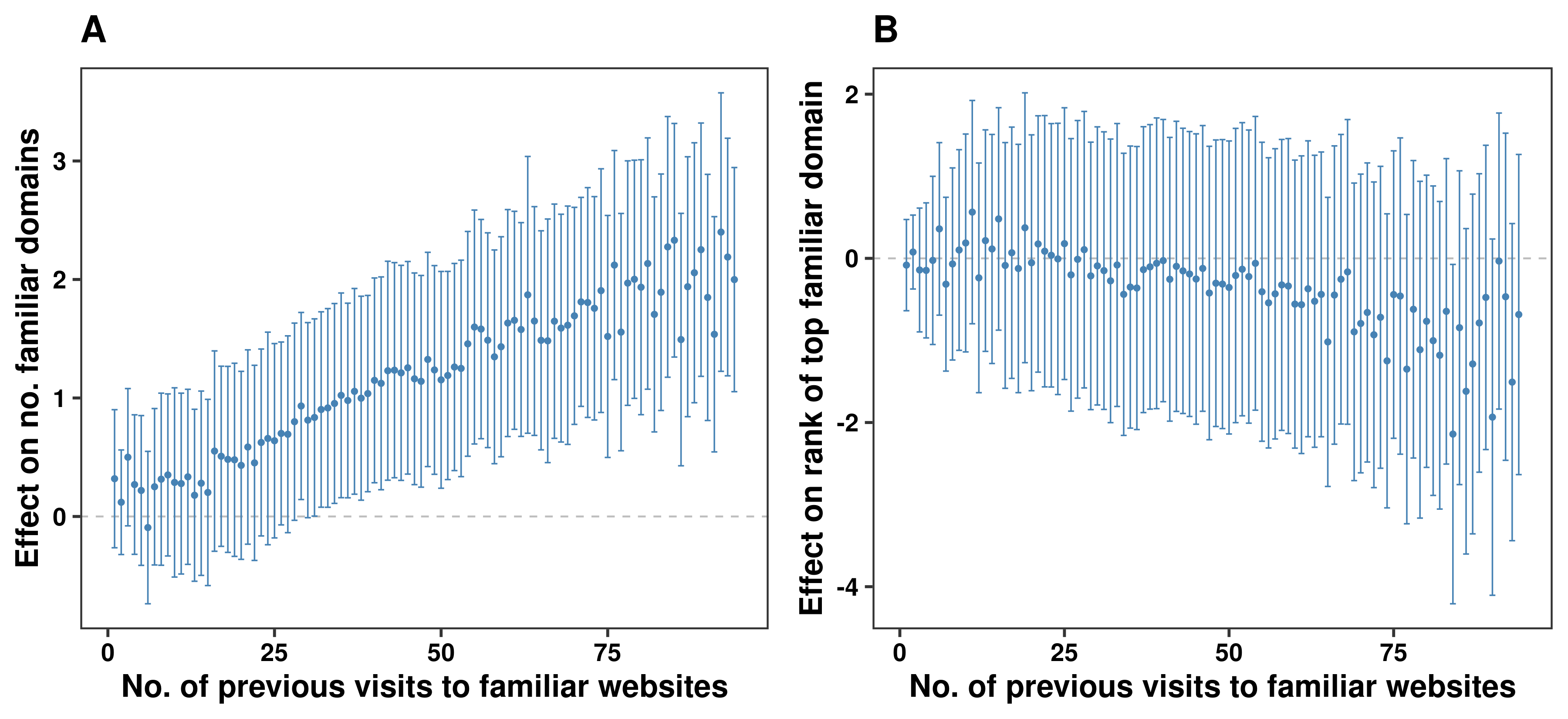}
\caption{Effects of the number of previous visits to familiar websites on the familiarity of the search results. The dependent variable is the number of familiar domains on the first search results page in panel A and the rank of the page's top familiar domain in panel B (where we restrict the sample to result pages that contain at least one of the corresponding user's familiar domains). Blue dots display marginal effects estimated from regressing the dependent variables on a set of indicator variables, one for each value of \emph{No.\ of previous visits to familiar websites} (with 0 visits as reference category), accounting for date-of-search, search-term, and browser-language fixed effects.  
Blue bars indicate 95\% confidence intervals based on standard errors three-way clustered by date of search, search term, and user.  
Extreme outliers (with values in the top 0.5\% of the number of previous visits) are excluded.
}
\label{fig:familiarity}
\end{figure}

To better understand Google Search's algorithmic personalization of search results, we test whether search results pages are more likely to contain the domains of websites that the user visited in the past. For each user and each election-related Google search, we thus count (i) how often a user has already visited websites from her personal set of  (partisan or non-partisan) domains and (ii) the number as well as the rank of such familiar domains on the first results page. 
Figure~\ref{fig:familiarity} shows in panel A that more previous visits to their favorite domains increases the number of these domains on the first result page -- even though the users are not logged into any account. This effect is even more pronounced for the top stories section (see \emph{Supplementary Material}, Figure~S9). Hence, taking non-personalized search results pages as the benchmark, we conclude that Google Search's algorithmic personalization prioritizes previously visited websites. 

Panel B of Figure~\ref{fig:familiarity} shows that more previous visits to familiar domains also lowers the rank of these domains, conditional on them appearing on the first result page. This effect, however, is comparatively weaker.

While personalization based on past visits exists, and users are set to visit partisan websites, we do not observe differences in the ideology of search results between left and right bots. This apparent inconsistency is due to popular and relatively centrist domains driving the observed increase in the number of familiar domains seen by users. Since domains on the fringes are extremely unlikely to occur on a search results page, we do not observe significant changes for these domains. Even if past visits increase the ranking of a highly partisan domain, the effect is not strong enough to be relevant when viewing the top search results. Since the domains used to set users' partisan identity, particularly on the right, are not nearly as popular as CNN, Fox News, or NY Times, we do not observe differences in search result ideology between partisan users. 

The fact that ideological differences exist for Democratic and Republican cities implies that location is weighted more heavily than past browsing behavior when personalizing search results. The finding that differences between cities are driven by local domains being shown to users despite having limited national appeal corroborates the focus on location for personalization.

\subsection*{Finding III: Search results do not generally prioritize websites featuring the user's ideology} 

Finally, we study whether Google Search prioritizes partisan websites beyond showing a user's favorite websites. In doing so, we focus primarily on ``new domains'' (defined as domains outside of the respective user's set of favorite domains). We first build a Search Result Ideology Score (SRIS) that measures the ideological leaning of organic search results to election-related queries based on five website ideology indices \cite{bakshy_etal2015,Robertson_et_al_2018,mitchell_etal_2014,budak_etal2016} 
(see Materials and Methods for details). This score ranges from -100 (most liberal) to 100 (most conservative). 
We then regress the SRIS on indicator variables for the users' partisan preferences and the prevalent partisanship in the city (with non-partisan users and purple cities as reference categories).  

\begin{figure}[tbh]
\centering
\includegraphics[width=11.4cm]{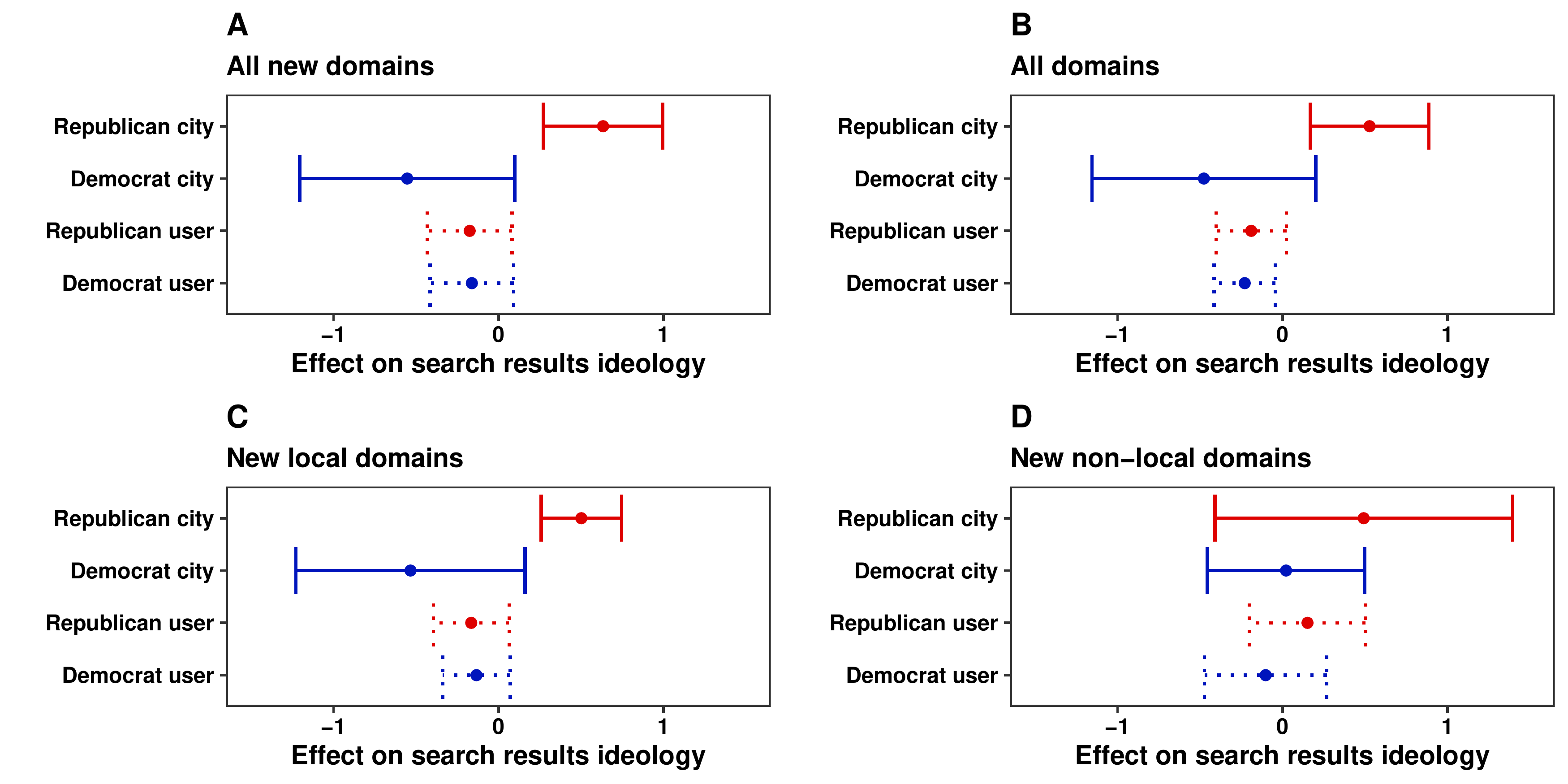}
\caption{Effects of partisanship and location on search result ideology of the set of domains indicated on top of each panel. Dots indicate marginal effects estimated from regressing the Search Result Ideology Score on indicator variables for the user's partisanship and their cities' partisanship (with non-partisan users and ``purple'' cities as reference categories), accounting for date-of-search, search-term, and browser-language fixed effects. The indicated 95\% confidence intervals are based on standard errors three-way clustered by date of search, search term, and user. Table~S6 in the \emph{Supplementary Material} shows the underlying regression estimates and table~S7 linear hypothesis tests.}
\label{fig:ideology}
\end{figure}

The results reported in panel A of Figure~\ref{fig:ideology} indicate that the users' partisanship does not play an important role, which suggests that Google Search does not show new partisan content based on partisan browsing and search histories. However, users in Democrat cities get more liberal/less conservative new content in their organic search results than users in Republican cities. 
In panel B, we redo the analysis using the SRIS based on all (rather than just new) domains. We again find that differences in the users' partisanship play no role, while differences in their locations' partisanship matter. These results are robust to alternative specifications and the use of individual website ideology indices, and it is similar for Democrat and Republican users (see Materials and Methods). Hence, taking non-personalized search results as the benchmark, we conclude that Google Search's algorithmic personalization prioritizes search results representing the locally prevalent ideology. 

In panels C and D, we disentangle the results reported in panel A by differentiating between new local domains and new non-local domains (see Materials and Methods for the coding of local domains). The results suggest that the prioritization of the locally prevalent ideology is driven primarily by local websites.

\section{Discussion}\label{sec:disc}
To summarize, we find considerable diversity across Google Search results in response to the same election-related queries (even though users are not logged into any account). More importantly, Google Search's algorithmic personalization prioritizes the users' favorite websites and (local) news sites. According to various independent measures, these sites are ideologically close to the locally dominant political party. Hence, compared to the benchmark of non-personalized search results, Google Search's algorithmic personalization users located in rather liberal cities tend to see systematically more liberal search results in comparison to users located in rather conservative cities (independent of their own political ideology). This, in turn, may imply that users are exposed to relatively few websites that contradict their own viewpoints. Given the well-documented preferences of individuals for reading and listening news that confirm their own beliefs, this feature of web search personalization may well be in the individuals' own narrow self-interest. However, it may not be in society's best interest if democracies work better when all citizens are exposed to political information from diverse viewpoints \cite{Sunstein_2001,finkel_etal2020}. 

Many scholars argue that political polarization has recently intensified in the United States  \cite{poole_rosenthal1984,abramowitz_saunders1998,Dimock_etal2014,iyengar_westwood_2015, iyengar_2019} and elsewhere \cite{mccoy_etal2018,hobolt_leeper_tilley_2020}. It is further argued that these developments have at least in part been the result of ``echo chambers'' and ``filter bubbles'' caused in one way or the other by the increasing importance of the world wide web \cite{dimaggio_etal2001,Sunstein_2001, Gentzkow_Shapiro_2011_QJE,bakshy_etal2015, Flaxman_et_al_2016,lelkes_etal2017, bail_etal2018,vosoughi_etal2018,Levy_2020,allcott_etal2020,finkel_etal2020,hosseinmardi_etal2021,waldrop_2021,watts_2021, cinelli_2021}. Moreover, it is argued that such a development could have far-reaching negative political and social consequences \cite{mccoy_etal2018,chen_rohla2018,finkel_etal2020, waldrop_2021}.
Importantly, our results cannot speak to the question about whether or not search engines -- as one key part of the world wide web -- have contributed to these undesirable social and political phenomena. We study the causal effects of algorithmic personalization on what users see in election-related search results and compare them to the benchmark of no search result personalization. Therefore, by design, our study cannot assess the overall effect of having search engines in contrast to having no search engines. However, it is remarkable that Google Search's algorithmic personalization may induce patterns of news consumption that are not too dissimilar from the patterns common prior to the ascent of the world wide web. In these times, people may have had a print subscription to a particular (likely local) newspaper or they may have gone to a news stand to buy newspapers they knew and that were locally well read. Nowadays, Google Search is proposing them websites they know and that feature the locally prevalent ideology in response to their political queries. Future research should study the interaction of informational segregation in online and offline news consumption and how conducive these two types of news consumption are to political polarization.

At a more general level, we believe that our approach, which combines synthetic users that build up browsing and search histories over several months with regular experiments in which all users use the same search term, is promising for future research. A similar approach could be successfully applied to investigate the effects and consequences of recommender systems on other web platforms, ranging from Amazon to Yahoo News or YouTube.

\section{Methods}\label{sec:methods}

\subsection{Technical implementation}
Our study's technical realization is based on a custom-made software designed to set up, configure, and control synthetic web users (`web bots') who emulate human user characteristics and device fingerprints (various technical characteristics of the hardware and software used to visit a website), and human-like browsing and search behavior. Figure~S2 in the \emph{Supplementary Material} illustrates the key components of the application's architecture.


At the lower level of the application, the ``runner instance'' handles and runs the users' browsing sessions, relying on a specialized remote web driver. The web browsing of our synthetic users is not recognizable as run by browser automation (unlike browsers automated by a standard web driver such as Selenium). Automated browser instances are then linked to two components that help us emulate unique human user characteristics: a fingerprint manager (allowing the user to appear as if it uses a specific operating system and hardware setting), and a residential proxy service (allowing the user to access the web through a residential IP address in a specific US city). Residential proxies use IP addresses provided by common Internet service providers and are thus not distinguishable from an IP address of a typical web user (unlike common proxy servers and VPN services, which are run in data centers and are easily identifiable as such based on their data center IP addresses). All web traffic issued and received by the users is recorded in HAR files (commonly used for web archiving). This allows us to see what a given user did at any point in time and what the user was exposed to on Google Search and other websites. 

\subsection{The population of synthetic users}


Our 150 unique user profiles are designed to reflect extended browser/device-fingerprint characteristics. This serves two purposes in our study design. First, from the perspective of websites visited by our synthetic users, these users look perfectly realistic as they appear to have common web browsers, screen resolutions, sound devices, and so on. 
Second, we can assure that each of the synthetic users has a unique fingerprint, such that it could be uniquely identified and tracked on that basis (in addition to or instead of relying on tracking cookies). 


We verify the synthetic users' appearance from the visited websites' perspective in two ways. 
First, we use security testing tools to verify that the synthetic users' device and network connection correspond to the intended configuration and that their fingerprints are unique (such that each user can be successfully tracked and uniquely identified). Second, we use an independent real-time geolocation service to verify the location of the residential proxy servers for each user's browsing sessions. Both verification tests confirm the validity of our approach (see \emph{Supplementary Material}, Section~C, for details).

\subsection{Synthetic users' browsing behavior}
The users' favorite non-partisan websites are randomly selected from the 100 most popular US websites according to \cite{ahrefs_2020}. Their favorite partisan websites are randomly chosen from the top 100 partisan websites used by Democrat (Republican) supporters. We determine the top 100 Democrat (Republican) partisan websites based on a collection of over 140M partisan tweets issued during the 2018 mid-term elections \cite{wrubel_etal2019}. We first extract and parse all URLs appearing in pro-Democrats tweets and in pro-Republican tweets (additionally filtering out URLs containing the domain twitter.com, abbreviated URLs, and domains no longer in use by October 2020). Following the idea of detecting partisan phrases in text \cite{Gentzkow_Shapiro_2010}, we then compute the ``partisanship'' of each domain/website $i$ as 
\begin{equation}
     \chi^2_{i}=\frac{(f_{ir}f_{\sim id}-f_{id}f_{\sim ir})^2}{(f_{ir}+f_{id})(f_{ir}+f_{\sim ir})(f_{ir}+f_{\sim ir})(f_{\sim ir}+f_{\sim id})},
\end{equation}
where $f_{id}$ ($f_{ir}$) denote the total number of times domain $i$ is referred to in a tweet by Democrat (Republican) supporters, and $f_{\sim id}$ ($f_{\sim ir}$) denote the total number of times a domain other than $i$ is referred to in a tweet by Democrat (Republican) supporters. A higher $\chi^2_{i}$ value indicates that $i$ is predominantly referred to by supporters of one of the two parties. Given $f_{id},\, f_{ir}$ and $\chi^2_i$ we can therefore select the 100 most partisan domains used by Democrat supporters and the 100 most partisan domains used by Republican supporters. Table~S2 in the \emph{Supplementary Material} shows the top 20 most partisan domains predominantly used by Democrat or Republican supporters, respectively.


\subsection{Synthetic users' search behavior}

Each user issues 1--2 non-partisan Google searches per day, such as unit conversions or very common search terms (e.g., ``ups tracker''). In addition, each Democrat (Republican) user gets assigned 60 liberal (conservative) search terms, randomly chosen from a list of liberal (conservative) search terms, and their set of partisan domains.

The lists of liberal (conservative) search terms consist of a large set of phrases used in national US politics that are most indicative of a certain ideology and relatively frequently used as search terms on Google Search. We build these lists in four steps: 
First, we collect data on all phrases used by Members of Congress (MoC) in tweets and congressional speeches during the 116th US Congress (see \emph{Supplementary Material}, Section~D for details). Following the procedure suggested in \cite{Gentzkow_Shapiro_2010}, we compute the ``partisan loading'' of all the (stemmed) bigrams in these data and select the 500 most partisan bigrams. These 500 partisan bigrams map to roughly 1,400 complete phrases.
Second, we use data on each MoC's ideological position from Voteview (https://voteview.com/data) to map our 500 bigrams to an ideology scale from -1 (clearly liberal) to 1 (clearly conservative), again closely following \cite{Gentzkow_Shapiro_2010}. 
Third, for each of the partisan phrases, we use Google Trends to verify whether, when, and where it was used as a search term on Google Search. We only keep the (unstemmed) partisan bigrams used as search terms on at least 50 days and in at least 10 US states since 2016. 
Lastly, we extend the compiled set of partisan search terms with search queries that, according to Google Trends, are related to (or substitute for) our partisan search terms. 

As an example, consider the (stemmed) bigram ``clean energi'', which we identify as one of the most partisan bigrams and a clearly liberal term. From all the unstemmed bigrams mapping to ``clean energi'', ``clean energy'' is the one most frequently used as search term. Finally, based on the queries related to ``clean energy'', we can map ``clean energi'' to the ``synonymous'' search terms ``solar energy'', ``clean renewable energy'', ``renewable energy'', and ``clean energy''. Out of the roughly 1,200 partisan search terms identified in this way, we select the 400 most clearly partisan search terms, label the conservative (liberal) ones as Republican (Democrat), and use them as the basis for the users' search vocabulary.

To validate our final selection of partisan search terms, we compute the relative frequency of the use of Republican as opposed to Democrat search terms for each US state. We document that these relative frequencies are positively correlated with the Republican vote shares across US states. The raw correlation coefficient is 0.49 (see \emph{Supplementary Material}, Section~E and Figure~S5, for details).

\subsection{Quantifying the similarity of search results pages}

For each first search results page of each pair of synthetic users who issued an identical election-related Google search, we compute the well-known Jaccard Index \cite{jaccard_1901} and the extrapolated Rank-Biased Overlap (RBO) \cite{webber_etal2010}. Formally, the extrapolated RBO of search result lists $S$ and $T$ is defined as
%
\begin{equation}
    RBO_{EXT}(S,T,p,k)=\frac{X_{k}}{k}\times p^{k} + \frac{1-p}{p} \sum_{d=1}^{k}{\frac{X_{d}}{d}\times p^{d}},
\end{equation}
where $k$ is the number of (observed) domains in the lists, $d$ is the depth or rank of the domains compared, and $p$ is the comparison ``persistence'' indicating how much weight is given to lower-ranked domains as opposed to higher-ranked domains. 
In order to weight search results in a similar manner to human users, we compute the average click through rate of results by their rank on the result page \cite{CTR}. 
For simplicity, we compute the weighting for a representative result page with eight visible items (as eight is the most common number of organic search results on the first search results page) in the categories ``Law, Government \& Politics'' and ``Weather, News \& Information.'' We find an average click through rate of 70.6\% across these categories, which implies a value for $p$ of around 0.9. 



\subsection{Quantifying search result familiarity}

To quantify how ``familiar'' the content of search results pages is to synthetic users, we count the number as well as the rank of domains from the users' personal set of favorite websites on the first results page. In addition, for each day of the observation period, we count how often a user has already visited a website from her set of favorite domains that ever occur in the election-related search results of any user. 

\subsection{Quantifying search result ideology}

To test whether user characteristics have an effect on the aggregate partisan leaning of a search results page beyond the re-occurrence of domains pointing to familiar websites, we employ a set of measures that index websites on a liberal-conservative scale. Specifically, we compute the Search Result Ideology Score (SRIS) from the $k$ domains listed on the first search results page as
\begin{equation}
\label{eq:ideology}
        SRIS = \frac{\sum_{d=1}^{k} \pi_d w_{d}}{\sum_{d=1}^{k} w_{d}},
\end{equation}
where $d$ is the rank of the domain, $\pi_d$ a proxy for the ideological leaning of the website behind this domain $d$, and 
\begin{equation}
\label{eq:familiarity_weight}
    w_{d} = (1-p)\times p^{d-1}
\end{equation}
its weight given its rank. As above, we set $p=0.9$ to make the weighting consistent with the empirical CTR.

We rely on five existing website ideology indices to compute $\pi_{d}$ \cite{bakshy_etal2015,Robertson_et_al_2018,mitchell_etal_2014,budak_etal2016}. These indices are based on very different methodologies, ranging from surveys and expert reviews \cite{mitchell_etal_2014} and the ratings of websites by human raters on MTurk \cite{Robertson_et_al_2018} to text analysis of (online) news outlets \cite{budak_etal2016} and the analysis of sharing behavior on Facebook and Twitter by liberal or conservative users \cite{bakshy_etal2015,Robertson_et_al_2018} (see \emph{Supplementary Material}, Section~G for more details on these indices). To make these indices comparable, we re-scale them all to a liberal-conservative scale in $[-100,100]$. For each website domain listed in the first search results page, we then compute $\pi_{d}$ by averaging all available website ideology indices (setting non-available index values to 0). We then use this average ideology index $\pi_{d}$ to compute SRIS from the $k$ domains listed on the first search results page. Consequently, a search results page is only assigned a clearly liberal or clearly conservative value if several website ideology indices assign a similar ideology score to some of the websites.

\subsection{Coding of local websites}

The coding of local websites underlying panels C and D of Figure \ref{fig:ideology} is based on two components (see \emph{Supplementary Material}, Section H, for details). First, for each domain, we verify whether a website is listed in a Media Cloud \cite{Roberts_etal2021} US ``States \& Local'' collection (largely about a particular state or a particular locality/city), but not in a US ``national'' collection (largely about the US as a whole).
Second, as the data by \cite{Roberts_etal2021} was generated at different points in time for different websites, we independently validate this categorization. For this purpose, we visit all websites initially coded as local and gather text describing the website and its purpose from the underlying source code. If a website's description also indicates a national scope (or neither a local nor a broader scope), we check the website for such information manually. Finally, we also manually check all discrepancies (coded as local according to Media Cloud, but not according to our validation) by visiting the corresponding website.


\subsection{Robustness of main findings}

We check the robustness of the main finding on search result familiarity reported in  Fig.~\ref{fig:familiarity} in several ways. For simplicity, we regress the number of familiar websites occurring in the search results (and related measures) on the number of previous visits to familiar websites. Table~S4 in the \emph{Supplementary Material} shows that the results are robust to varying fixed-effects specifications, varying cluster-robust standard error estimations, the inclusion of a linear time trend, the inclusion of the number of previous searches of familiar domains (e.g., a user typing nytimes in the Google search bar instead of directly visiting www.nytimes.com), as well as to alternative codings of the dependent variable. Table~S5 in the \emph{Supplementary Material} shows that the results are even somewhat more pronounced for the search results' top stories section. 

We also check the robustness of our main findings on search results ideology reported in Figure~\ref{fig:ideology}. Table~S6 in the \emph{Supplementary Material} shows the regression output behind Figure~\ref{fig:ideology} (see columns 1-4) as well as additional specifications in which we replace our average SRIS with the scores based on the individual website ideology indices (see columns 5-9). While, not surprisingly, the results vary from index to index, the overall picture is qualitatively consistent with the baseline specification based on the average SRIS. This reaffirms that the results of the baseline specification are not driven by one particular website ideology index. In addition, Table~S7 in the \emph{Supplementary Material} shows linear hypothesis tests of whether the effects are the same for Democrat vs.\ Republican users and for users from Democrat vs.\ Republican cities. Table~S8 in the \emph{Supplementary Material} shows alternative specifications where we regress the average SRIS on the average ideology score of the users' previously visited favorite websites (rather than indicator variables for the user's partisanship) and the city-level share of Republican votes (rather than indicator variables for Democrat and Republican cities). The results are consistent with those reported in Figure~\ref{fig:ideology}. 
Finally, Figure~S10 in the \emph{Supplementary Material} shows that both Democrat and Republican users experience a similar prioritization of the locally dominant ideology. 

\clearpage

\bibliography{sn-bibliography}


\end{document}


\title{\textbf{Supplementary Material:}\\Personalization of Web Search During the 2020 US Elections}
\author{Ulrich Matter, Roland Hodler, Johannes Ladwig}


\maketitle


\renewcommand\thesection{\Alph{section}}
\onehalfspacing

\section{Study Design and Context}

This part complements our discussion of previous contributions and the description of our study design in the research article. Specifically, it discusses the challenges related to measuring effects of the algorithmic personalization of search results and how our study design and the setting of our study address these challenges.

\subsection{Challenges to the measurement of search result personalization}

The design of this study is aimed at meeting two key challenges regarding the measurement and identification of the effect of the search engines' algorithmic personalization on systematic differences in political search results. First, the personalization of search results is only a second order concern for major search engines like Google Search. The first order concern is that search results are objectively relevant, given a search term.\footnote{The relative importance of webpage characteristics vs. user characteristics is prominent in the information retrieval literature. For example, in \cite{buttcher2016information} (p.3) the task of information retrieval in a web search context is described as follows: 
``From [...] millions of possible pages, a search engine's ranking algorithm selects the top-ranked pages based on a variety of features, including the content and structure of the pages (e.g., their titles), their relationship with other pages (e.g., the hyperlinks between them), and the content and structure of the Web as a whole. For some queries, characteristics of the user such as her geographic location or past searching behavior may also play a role.''} Therefore, the broader context of studying the effects of web search personalization on informational segregation is of substantial relevance -- not just for matters of external validity but also regarding identification. Most importantly, personalization effects cannot be measured (and thus the relevant factors driving personalization cannot be identified) if other aspects of the search algorithm and factors outside of the search engine fully dominate the results.\footnote{Other aspects that are considered highly relevant in the information retrieval literature are the relevance of a webpage's content for the search term used in the query, as well as the webpage's prominence in the web. The latter is usually measured with the page rank algorithm \cite{page_etal1999}.} For example, suppose a specific topic is exclusively covered by cnn.com. Even if user data suggest a certain user clearly prefers foxnews.com over cnn.com, the user would only see cnn.com-webpages listed in her search results when querying Google Search regarding this specific topic (since there would be no objectively relevant alternative). Concluding from this that Google Search does not tailor search results to its users' preferences would obviously be problematic. Hence, it is important that the search terms used to test for such personalization effects fit a selection of prominent pages of websites across the political spectrum. Whether that is the case in a specific setting is ex ante very difficult to know for the researchers, because the web and therefore the relevant search terms are constantly changing. In consequence, there is always a reasonable concern that estimated personalization effects are biased towards zero.

Second, the users employed in the study need to be clearly distinguishable for the search engine in the dimensions that matter for the research question  (i.e., in the context of this study, with regard to the user's browsing habits, their political preferences, and their location). Yet, the users' behavior should also be realistic and approximately representative for a relevant part of the overall population.
On the one hand, users who are not clearly different in the relevant dimensions are unlikely to see very different results on the first search results page (even if the context of the study is carefully chosen and the users are representative of a relevant part of the population, as, for example, in \cite{Krafft_2019}). In this case, there would again be a concern that estimated personalization effects are biased towards zero. 
On the other hand, it can also be problematic to employ users with rather extreme behavior (i.e., users who intensely and exclusively consume information from one or the other extreme of the dimension of interest, as in \cite{le_etal2019}), because it may reduce the results' external validity.

\subsection{Ideal testing ground and synthetic users}

In order to meet the challenges outlined above, we study the algorithmic personalization of search results in the setting of the 2020 US Presidential Elections and use an approach complementary to the current literature: a long-term study based on synthetic internet users. This setting allows us to address the issue of downward bias as well as the issue of external validity.

First, we think that the 2020 US Presidential Elections were an ideal testing ground to study whether and how the algorithmic personalization can lead to systematic differences in political search results. Generally, US presidential elections generate a lot of media attention across the political spectrum in the United States. In addition, the 2020 Presidential Elections were expected to be very tight in some key swing states and the outcome was controversial and contested due to allegations by then-President Donald J.\ Trump (which created even more media attention). Finally, a substantial part of the electorate was fairly polarized and could be clearly characterized as either Democrat or Republican (in terms of personal media consumption, group ideology, and geography). That is, on the one hand, there was plenty of topical information provided from various prominent web sources across the political spectrum, and, on the other hand, a large share of the population was searching for this information on Google Search. By testing personalization based on the most frequent election-related search terms used on Google Search (according to Google Trends), we address both the issue of external validity and also substantially reduce the risk of a downward bias due to a lack of objectively relevant and diverse search results.

Second, by combining the advantages of two seminal former approaches (short-run controlled experiments \cite{hannak_2013, Hannak_et_al_2017, le_etal2019} and studies based on `data donations' \cite{Robertson_et_al_2018, Krafft_2019}), we can strike a fair balance regarding the trade-off between distinguishable user profiles (generating the variance necessary to identify personalization effects) and external validity (avoiding very intense/extreme information consumption patterns over relatively short periods of time). As we have full control over the (randomized) assignment of user characteristics, there is no concern about unobservable similarities between users. As we have full control of how users build their browsing and search histories over several months, spurious results attributable to a snap-shot of the search algorithm and the momentary state of the web are very unlikely.

\section{Technical Implementation}

This part complements the short summary of the technical implementation in the research article by discussing how we verify our implementation. 

\subsection{Verification of fingerprinting and geolocations}

We assess the users' appearance with a set of security testing tools provided by BrowserLeaks (\url{www.browserleaks.com}). To this end, we extract and parse the reports generated by BrowserLeaks for each of our synthetic users and verify a) whether the synthetic users' device and network connection is consistent with the intended configuration, and b) whether our users' fingerprints are in fact all unique (and hence, our users can be successfully tracked and uniquely identified based on them). Figure~\ref{fig:browserleaks} shows a sample screenshot taken from such a verification test.


Second, we monitor the residential proxy servers via an independent real-time geolocation service at the beginning and at the end of each user's browsing and search sessions. Figure~\ref{fig:deviations_map} shows the 3rd-party verified geolocations. Blue crosses mark the verified coordinates of users when browsing and searching (all browsing sessions of all users are included in the plot), orange circles highlight the official city coordinates to which synthetic users were assigned during the study. Synthetic users were generally recognized as being located in their assigned city, with very few exceptions. We see that very few of the thousands of geocoded browsing sessions (blue crosses) are outside of the orange circles.

\section{Input Data and Synthetic User Configuration}

Complementing the discussion of the configuration of search and browsing behavior in the research article, this part explains in more detail how we select and assign partisan search terms to synthetic users, and how we validate our selection of partisan search terms.

\subsection{Partisan search terms}

In the research article, we sketch how we determine the partisan search terms that the synthetic users use on Google Search in order to find and consume partisan content. In this section, we provide more detailed information on how we generate empirically reasonable lists of liberal and conservative search terms. We do so because the partisan search terms are important, as they allow Google Search to learn about our users' partisan preferences (beyond what is reflected in their direct visits to their favorite domains). 

The aim is to collect a large set of phrases used in national US politics that are most indicative of a certain ideology and at the same time relatively frequently used as search terms on Google Search. We proceed in four steps.

First, following the procedure suggested in \cite{Gentzkow_Shapiro_2010}, we compute the `partisan loading' of phrases (bigrams) used in national US politics. To this end, we collect data on all phrases used by Members of Congress (MoC) in tweets and congressional speeches during the 116th US Congress (the period relevant for our study). Speech data is collected from the Congressional Record provided in digital form by the Library of Congress (see \url{https://www.congress.gov/congressional-record}); tweets are collected from the MoCs' twitter feeds. We only use those bigrams considered `valid' in the sense of \cite{gentzkow_etal2019} (i.e., procedural bigrams, non-speech-related bigrams from the Congressional Record, etc. are removed). For all processing of text data described in this subsection, we use the stemmed bigrams. We then compute the `partisanship' of each (stemmed) phrase/bigram $p$ as 
\begin{equation}
     \chi^2_{p}=\frac{(f_{pr}f_{\sim pd}-f_{pd}f_{\sim pr})^2}{(f_{pr}+f_{pd})(f_{pr}+f_{\sim pr})(f_{pr}+f_{\sim pr})(f_{\sim pr}+f_{\sim pd})},
\end{equation}
where $f_{pr}$ and $f_{pd}$ denote the total number of times bigram $p$ is used by Republicans and Democrats, respectively, and $f_{\sim pr}$ ($f_{\sim pd}$) denote the total number of times a bigram that is not bigram $p$ is used by Republicans (Democrats). A higher $\chi^2_{p}$ value indicates that $p$ is predominantly used by members of one of the two parties. For the next steps, we select the 500 most partisan phrases.
As an illustration, Table~\ref{tab:partisanbigrams} shows the 30 most partisan phrases predominantly used by Democrats and Republicans, respectively.


Second, we want to map these 500 bigrams to an ideology scale from -1 (clearly Democrat/liberal) to 1 (clearly Republican/conservative). To do so, we collect data on each MoC's ideological position from Voteview.\footnote{Voteview (\url{https://voteview.com/data}) provides estimates of MoCs' ideological positions on a scale from -1 (very liberal) to 1 (very conservative). The ideological positions (so-called DW-Nominate scores) are inferred from roll call records, using the scaling method suggested by Poole and Rosenthal \cite{poole_rosenthal1985}.} We denote their DW-Nominate score for MoC $c$ by $\pi_{c}$.
We next compute the relative frequency with which each MoC $c$ uses a given phrase/bigram $p$: $\tilde{f}_{pc}={f}_{pc}/\sum_{p=1}^{P}f_{pc}$. Again, closely following \cite{Gentzkow_Shapiro_2010}, we regress $\tilde{f}_{pc}$ on $\pi_{c}$ for each bigram $p$, which gives us intercepts $\alpha_p$ and slope coefficients $\beta_p$. A positive $\beta_p$ means the more often a congressperson uses $p$ (relative to other terms), the more Republican/conservative she is. $\beta_p$ thus indicates bigram $p$'s location on the liberal-conservative scale. In the same vein, we interpret $SE(\beta_p)$ as an indication of whether bigram $p$'s position on the liberal-conservative scale is more or less precisely measured.\footnote{For example, a given phrase might be used often by rather moderate Republicans who are ideologically not so far from rather centrist Democrats. Now suppose these moderate Democrats and Republicans have relatively close but nevertheless clearly distinct positions on the liberal-conservative scale (according to their voting behavior, measured by DW-Nominate scores). It could well be that a given phrase is used only slightly more often by moderate liberals than by moderate conservatives and the $\beta_p$ is not statistically significantly different from 0. In such cases $SE(\beta_p)$ helps us to take into consideration whether a phrase is clearly indicative of a MoC's position on the liberal-conservative scale.} That is, we interpret a bigram $p$ with a positive and large t-value of $\beta_p$ as `clearly conservative' and a bigram $p$ with a large but negative t-value of $\beta_p$ as clearly liberal. Finally, we re-scale the t-values of $\beta_p$ to $[-1,1]$.

Third, we check whether the most partisan bigrams identified are actually used as search terms in Google Search. For each complete phrase matching one of the stemmed bigrams $p$ we verify whether, when, and in which region it was used as a search term in Google. We query the Google Trends platform (\url{https://trends.google.com}) for each of the completed (unstemmed) most partisan bigrams.\footnote{The 500 most partisan bigrams map to roughly 1,400 complete phrases. For example, the stemmed bigram `climat chang' maps to the complete phrases `climate change', `climate changing', `climate changes', `climate changed', `climatic change'. Per stemmed bigram, we only keep the corresponding unstemmed partisan bigram that is most frequently used as a search term in Google Search.} We then verify how often (in relative terms) each of the remaining unstemmed bigrams are in fact used as search terms in Google Search. Relative search term frequencies are provided by Google Trends. However, the relative frequencies are scaled to an $[0,100]$ interval, where the value $100$ indicates the search term with the highest relative frequency of a maximum of five search terms in a given time frame and region/state. The usage frequencies of search terms are thus always expressed relative to each other and relative to the region and time. As the Google Trends platform only allows comparisons of five search terms at a time, we query the search term frequencies in batches of four partisan bigrams and add to each set of four bigrams the search term ``carbon free'' (a frequently and constantly used search term across the US). This way, all search term frequencies are relative to the same reference frequency (both across states and over time). As potential search terms, we only keep the bigrams that are used on at least 50 days and in at least 10 US states since 2016 as potential search terms.


Fourth, we extend the compiled set of partisan search terms with search queries that, according to Google Trends, are related to (substitutes for) our partisan search terms. To this end, we use information (provided by Google Trends) which relates other queries to a given search query (as long as the latter is used rather often).\footnote{Specifically, ``[u]sers searching for [this] term also searched for these [related] queries'' (\url{https://trends.google.com/trends/}.} We use the `related queries' information to select for each partisan search term related queries that are searched for at least 90 percent as often as the corresponding original search term (using Google Trends' 
`Top' metric). We can think of these related queries as alternative formulations/synonyms of the original partisan search terms. Following up on the example from above, the stemmed bigram ``clean energi'' is identified as one of the most partisan bigrams and is clearly identified as a typically liberal term (with a $\beta_p$ t-value of $-6.44$, re-scaled to $\pi_p = -0.477$). From all the unstemmed bigrams mapping to ``clean energi'', ``clean energy'' is the one most frequently used as search term. Finally, based on the queries related to ``clean energy'', we can map ``clean energi'' to the `synonymous' search terms ``solar energy'', ``clean renewable energy'', ``renewable energy'', and ``clean energy''.


Out of the roughly 1,200 partisan search terms identified in this way, we select the 400 most clearly partisan search terms, label the conservative (liberal) ones as Republican (Democrat) search terms, and use them as the basis for the users' search vocabulary.\footnote{Specifically, we select the top 400 cases with an absolute value of $\pi_s$ greater than 0.5, and label search terms with negative $\beta_p$ values as Democrat and those with positive values as Republican.} We randomly assign to each Republican (Democrat) user a set of 50 Republican (Democrat) search terms. In addition, we assign each Republican (Democrat) synthetic user a set of 10 highly partisan (and election-related) hashtags to be used as additional search terms (collected from \url{http://best-hashtags.com/hashtag/republican/} and \url{http://best-hashtags.com/hashtag/democrat/}). Finally, the synthetic users are also configured in such a way as to directly use the domain names of their respective ten most favorite partisan websites as search terms (instead of typing the entire domain into the browser bar). The main purpose of configuring partisan users to search for partisan content in accordance with their favored party is to develop a behavioral pattern that can be used by Google Search to personalize search results (either based on specific user preferences or partisan group affiliation). 

\subsection{Validation of partisan search terms}

We validate our final selection of partisan search terms as follows. We denote $\Bar{f}_{ps}$ the average relative frequency with which a given phrase/bigram $p$ is used for Google queries from computers located in state $s$. This information is again provided by Google Trends and is defined as 
\begin{equation}
    \Bar{f}_{ps}=\frac{\sum_{t}^{T}[\frac{(f_{pst}/f_{rst})}{max(f_{pst}/f_{rst})}]\times 100}{T},
\end{equation}
where $T$ is the total number of days $t$ on which the search term frequencies were recorded and $f_{rst}$ is the query frequency of the reference term, in our case, ``carbon free''. First, focusing exclusively on the subset of search terms labelled as Republican, we then approximate the Republican share of the googling population in state $s$ with the relative Republican search volume in a given state $s$: 
\begin{equation}
RepSearches_{s}=\frac{{\sum_{p=1}^{P}{(\Bar{f}_{ps})}}}{P\times100}.    
\end{equation}
$RepSearches_{s}$ can range from 0 (none of the Republican search terms were used in $s$) to 1 (every Republican search term was used most in $s$). Then, we analogously compute $DemSearches_{s}$ based on the subset of search terms labelled as Democrat. Finally we compute the `net Republican search volume' per state as $RepSearches-DemSearches$. 
We then compare the net Republican search volume per state with the net share of Republican votes in the 2020 US Presidential Elections. Figure~\ref{fig:searchterms_validation} shows the result of this comparison. The two measures are remarkably positively correlated (the raw correlation coefficient is 0.49). This suggests that a higher share of Republican voters in a given state tends to be reflected in the Google search behavior of that state's population. In the aggregate, Google Search users in a more conservative state tend to use conservatively rather than liberally slanted search terms. Moreover, our validation results at the aggregate level are consistent with recent experimental evidence on partisan information seeking at the individual level \cite{peterson_shanto2021}. 
Methods
\section{Analysis of Search Results}

This part presents additional material related to the findings shown in the research article. 
In our analyses we focus primarily on differences in the organic search results, which are the main component of a search results page, and secondarily on differences in the top stories component. While all search results pages contain organic search results, top stories do not occur in all search results pages. Figure~\ref{fig:google_serp} indicates the two components in a screenshot of a Google search results page.

\subsection{Quantifying familiarity with search results pages: robustness checks and additional results}

We present a series of alternative regression specifications to demonstrate the robustness of the main findings shown in Figure~3 of the research article. To keep the display of results simple, we impose a linear relationship between the number of previous visits of familiar domains and the number of familiar domains (on the first results page) in all of these specifications. (UsiMethodsng the same flexible dummy-variable specification as in the research article does not qualitatively change the results shown here.) Table~\ref{tab:serp_familiarity} shows the estimates based on the organic search results data. Columns 1-4 show the basic model with varying fixed-effects specifications as well as varying cluster-robust standard error estimations. In column 5 we account for the time elapsed since the beginning of the observation period (i.e., the beginning of our study). We see that our main finding is not driven by a time trend towards generally more familiar search results. In columns 6 and 7 we account for the number of previous searches of familiar domains (e.g., a user typing nytimes in the Google search bar instead of directly visiting www.nytimes.com). While regressing the number of familiar domains in election-related search results on the number of previous searches of familiar domains yields an effect in the expected direction (column 6), adding this number as a control to our baseline specification does not change our main finding qualitatively (column 7). Column 8 is identical to column 4, but we restrict the sample to search results that contain at least one of the domains familiar to the user. Finally, in columns 9 and 10, we replace the number of familiar domains with alternative dependent variables. Column 9 is identical to column 4, except that the dependent variable is a simple indicator equal to 1 if the search result contains at least one domain from user $i$'s set of familiar websites and 0 otherwise. In column 10, we restrict the sample to search results that contain at least one of the domains familiar to the user and regress the rank of the (highest-ranked) familiar domain on the number of previous visits to familiar websites. The estimate shows that, conditional on the search results containing at least one familiar domain, the more often a user has previously visited familiar websites, the higher up in the search results those websites' domains occur. A total of 75 previous visits to familiar websites increases the rank (decreases the numerical value of the rank) of the familiar domain occurring in the search results on average by one rank.

These results are even more pronounced for the search results top stories component. Figure~\ref{fig:familiarity_topstories} shows the effect estimates for the baseline models based on the top stories component (in analogy to Figure~3 in the research article) and Table~\ref{tab:serp_familiarity_topstories} shows the additional robustness checks based on the top stories component (in analogy to Table~\ref{tab:serp_familiarity}) 

\subsection{Quantifying the partisan leaning of search results: details on ideology indices and additional results}

We use the following website ideology indices (presented in alphabetical order) to compute the Search Results Ideology Score (SRIS) defined in equation~3 in the research article. 

\begin{itemize}
    
    \item \emph{Bakshy et al.} \cite{bakshy_etal2015} propose a partisan `alignment score' that indexes websites on a continuous scale from $-1$ (liberal) to $1$ (conservative) based on the relative frequency with which webpages of these websites are shared on Facebook by self-identified liberal or conservative Facebook users. 
    
    \item \emph{Budak et al.} \cite{budak_etal2016} propose a `partisanship score' based on the depiction of the Republican Party and the Democrat Party in political news articles. The score indexes (online) news outlets on a scale from $-1$ (left leaning) to $1$ (right leaning).
    
    
    \item \emph{MTurk Bias Score by Robertson et al. \cite{Robertson_et_al_2018} }: The authors use human raters on MTurk to code a subset of websites used in their main index (see below) on a five-point Likert scale from -1 (liberal) to 1 (conservative). 

    \item \emph{Pew Research Center} (\cite{mitchell_etal_2014} adapted in \cite{Robertson_et_al_2018}): Mitchell et al.\ from the Pew Research Center use a survey with several policy-related questions to map survey participants on a five-point scale from consistently liberal to consistently conservative and study which news outlets the respondents trust most. Based on these survey data, Robertson et al. \cite{Robertson_et_al_2018} create an index on a liberal-conservative scale from $-1$ to $1$, reflecting which online news outlets tend to be trusted by liberals/conservatives. We use this index as provided by \cite{Robertson_et_al_2018}.
    
    \item \emph{Robertson et al.} \cite{Robertson_et_al_2018} propose a `partisan audience bias score` based on registered (Democrat or Republican) voters' sharing of web domains on Twitter. The score scales from $-1$ (the domain of a website is exclusively shared by registered Democrats) to $1$ (the domain of a website is exclusively shared by registered Republicans). 
\end{itemize}

Table~\ref{tab:serp_ideology_notknown}, columns 1-4 presents the detailed estimates behind Figure~4 of the research article. These specifications differ in the set of domains used to compute SRIS. Columns 5-9 show estimates of the same specification as in column 1 (which includes all new domains) when replacing the average SRIS with the scores based on the individual website ideology indices. That is, we compute the SRIS for each ideology index separately and regress the resulting score on the same co-variates as in the main specification. While, not surprisingly, the results vary from index to index, the overall picture is qualitatively consistent with the baseline specification based on the aggregate index. 
Table~\ref{tab:serp_ideology_linh_notknown} shows the corresponding linear hypothesis tests of whether the effects are the same for Democrat vs.\ Republican users and for users from Democrat vs.\ Republican cities for all columns in Table~\ref{tab:serp_ideology_notknown}. 

Table~\ref{tab:serp_ideology_userideology} replaces the indicator variables for the users' partisanship by the average ideology score of the users' previously visited favorite websites and the indicator variables for Democrat and Republican cities by the city-level share of Republican votes. The results confirm that the partisanship of the users' location is more important than the partisanship of their search and browsing history.

Figure~\ref{fig:city_samplesplit} estimates the city partisanship effects separately for Democrat users and Republican users. 
While the Republican city effect tends to be more pronounced for Democrat users than Republican users, the general direction of the city partisanship effects are the same: Both Democrat and Republican users tend to see more Republican content in Republican cities and more Democrat content in Democrat cities. 

\subsection{Classifying domains as local and non-local}

The categorization of websites into local and non-local underlying panels C and D of Figure~4 is based on two components. First, for each domain found in the election-related search results, we fetch domain-level information from Media Cloud \cite{Roberts_etal2021}.\footnote{See the Media Cloud Source Manager under \url{https://sources.mediacloud.org} as well as the API endpoints for programmatic access here: \url{https://github.com/mediacloud/backend/blob/master/doc/api_2_0_spec/api_2_0_spec.md}.} Based on Media Cloud's data, we code a search result item as local if the corresponding domain is listed in one of Media Cloud's US ``States \& Local'' collections (largely about a particular state or a particular locality/city), but not in a US ``national'' collection (largely about the US as a whole). 
Second, as the original coding by \cite{Roberts_etal2021} was done at different points in time for different websites, we independently validate this categorization. The validation is based on a four step process. First, we visit all websites initially coded as local and gather text describing the website and its purpose from the underlying source code. All text contained in relevant metadata tags such as ``description'', ``site name'' and ``title'' is scraped from each domain. Second, we find matches between the gathered text, the domain itself and lists of US place names  \cite{Census_Bureau}\footnote{We omit places including but not limited to ``fox'', ``globe'', ``how'' and ``media'' to avoid false matches.}. Additionally terms such as ``local news" and ``local'' were coded as evidence for a local scope. Similarly, we find  matches to a list of terms indicating a non-local reach of the website. Such terms include ``USA'', ``national'' and ``international''. If matches occur exclusively with the list of places and local indicators, the domain is deemed to be local. Analogously, if there are exclusively matches to the list of non-local indicators, a domain is classified as such. In the event that there is overlap and the metadata tags include matches from only the local list or exceeds the number of non-local indicators by a factor of three or more, the website is classified as local. In a third step, we categorize a domain as non-local if it has a foreign top level domain and as local if it contains a US state abbreviation (e.g., ``atg.wa.gov''). The remaining, unclassified, domains either had no information on location contained in their metadata tags or exhibited both local and non-local indicators. We classify these domains manually by checking domains for clear references to cities, states, or places that were not spotted in the automated process; or by visiting the website manually. In doing so, any website that makes no reference to a specific location is classified as non-local. In a final step, we manually check all cases that were coded as local according to Media Cloud, but not according to our validation. We do so by visiting the corresponding website.

\clearpage
\section{Figures}


\begin{figure}[h]
\centering
\includegraphics[width=0.99\textwidth]{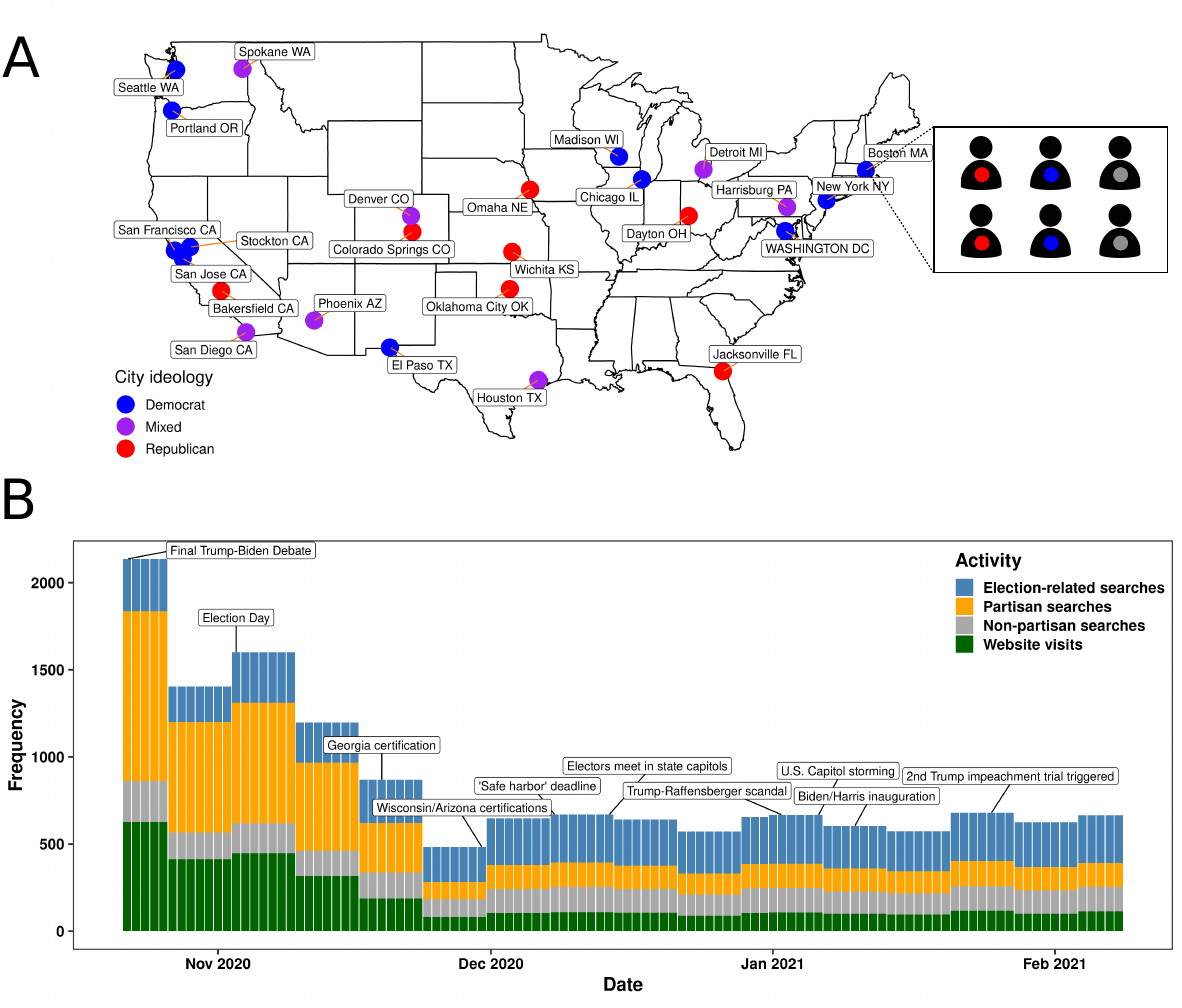}
\caption{Spatial distribution of users and timeline. Panel A shows the spatial distribution of the 150 synthetic users. The 25 cities include Democrat strongholds (blue), Republican strongholds (red), and `purple' cities, where neither party dominates. The city categorization is based on the Republican vote share in the 2016 US Presidential Elections (\emph{SI Appendix}, Table~S1). In each city, there are two Democrat (blue), two Republican (red), and two non-partisan users (gray), each with randomized differences in appearance and behavior. Panel B illustrates the timeline of our study and the users' weekly activities. Each bar represents the cumulative number of browsing and search tasks executed by the users (with the shading indicating the type of activity; weekly average per day). Labels point to relevant election-related events. }
\label{fig:design}
\end{figure}

\clearpage
\begin{figure}[h]
\begin{center}
   \includegraphics[width=0.75\textwidth]{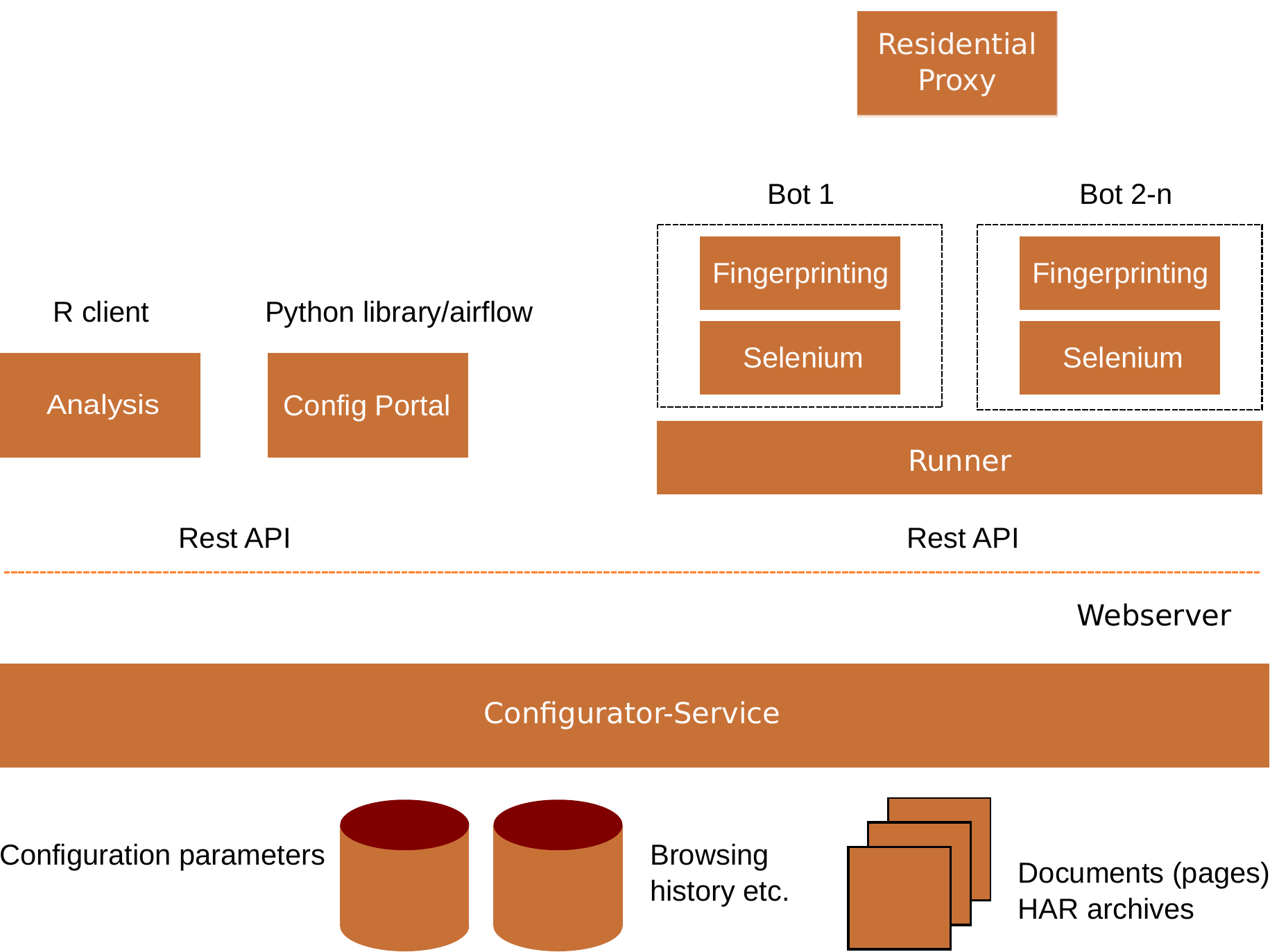}
\caption{Illustration of the basic software architecture. The software consists of three core components: The `runner' instance, which handles and runs the synthetic users' browsing and search sessions, the configurator-service (API), which stores and provides all user configurations, jobs, and recorded web traffic (in HAR files), as well as the configuration portal (a client library to set up and configure the user population).}
\label{fig:architecture} 
\end{center}
\end{figure}

\clearpage
\begin{figure}[p]
\begin{center}
   \includegraphics[width=0.95\textwidth]{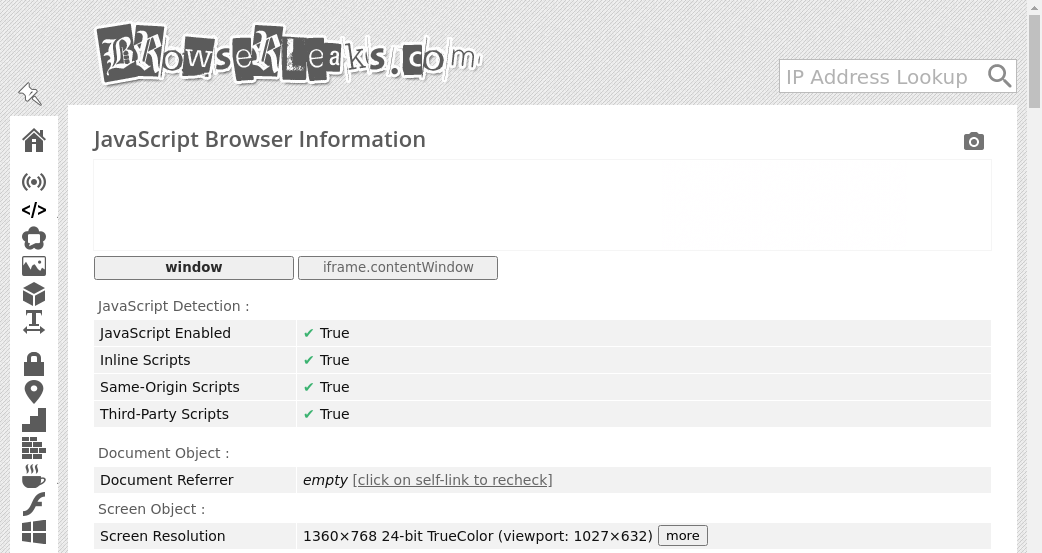}
\caption{Fingerprinting verification example. Sample screenshot showing the first few rows of a synthetic user's JavaScript browser profile on \url{www.browserleaks.com}. The example illustrates that the user's JavaScript settings are correctly recognizable and functional as well as that the synthetic user's (virtual) screen is properly recognized (screen properties recognition and canvas fingerprinting are some of many techniques used to identify and track users based on their device characteristics).}
\label{fig:browserleaks} 
\end{center}
\end{figure}

\clearpage
\begin{figure}[p]
\begin{center}
   \includegraphics[width=\textwidth]{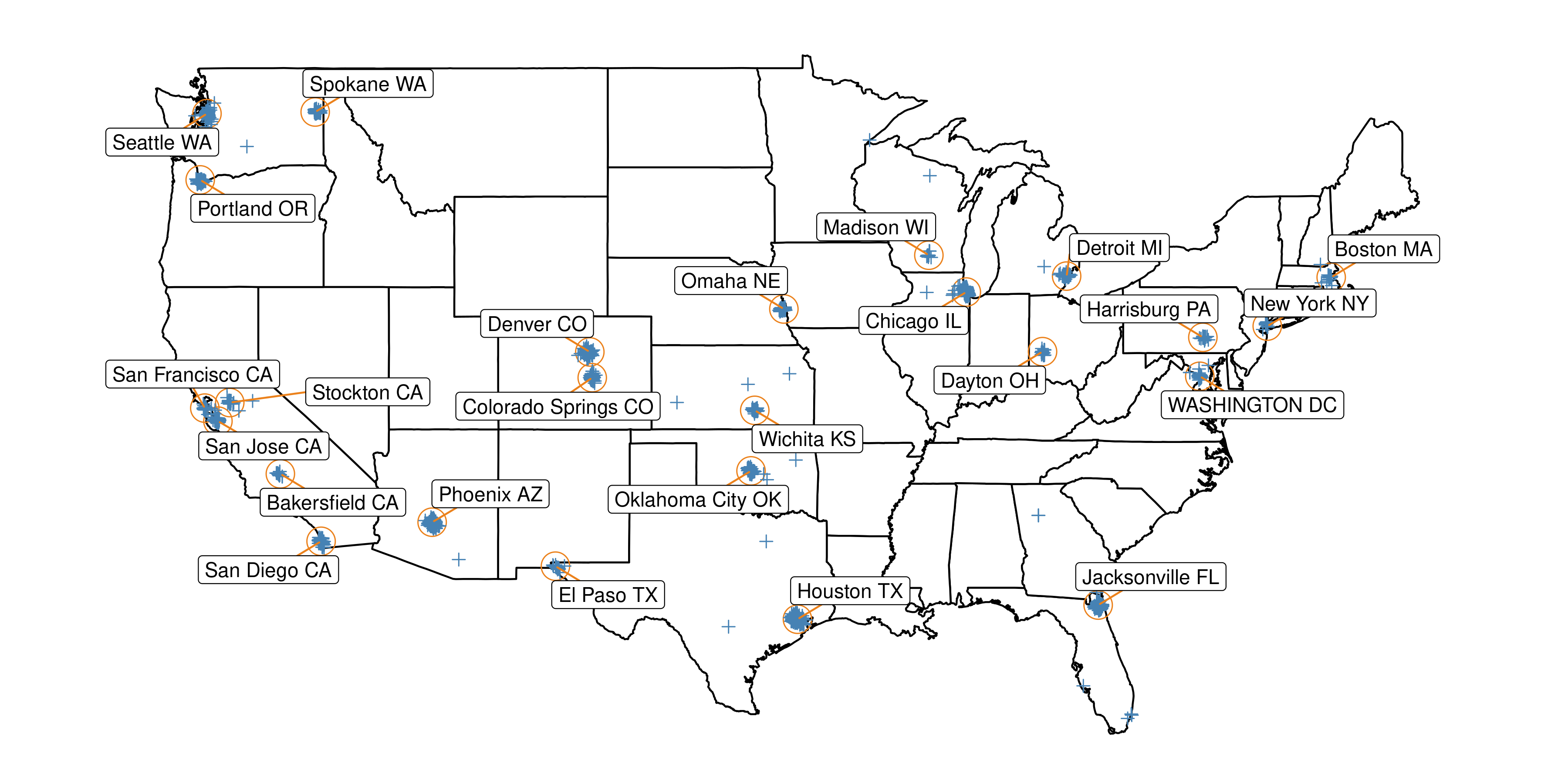}
\caption{Geolocation verification. Mapping of unique coordinates extracted during synthetic user geolocation verifications from all browsing sessions (based on the third-party real-time geolocation service IPStack). Blue crosses indicate the verified coordinates of users when browsing and searching (all browsing sessions of all users included), and orange circles highlight the official city coordinates to which synthetic users were assigned during the study.}
\label{fig:deviations_map} 
\end{center}
\end{figure}

\clearpage

\begin{figure}[p]
\begin{center}
   \includegraphics[width=0.75\textwidth]{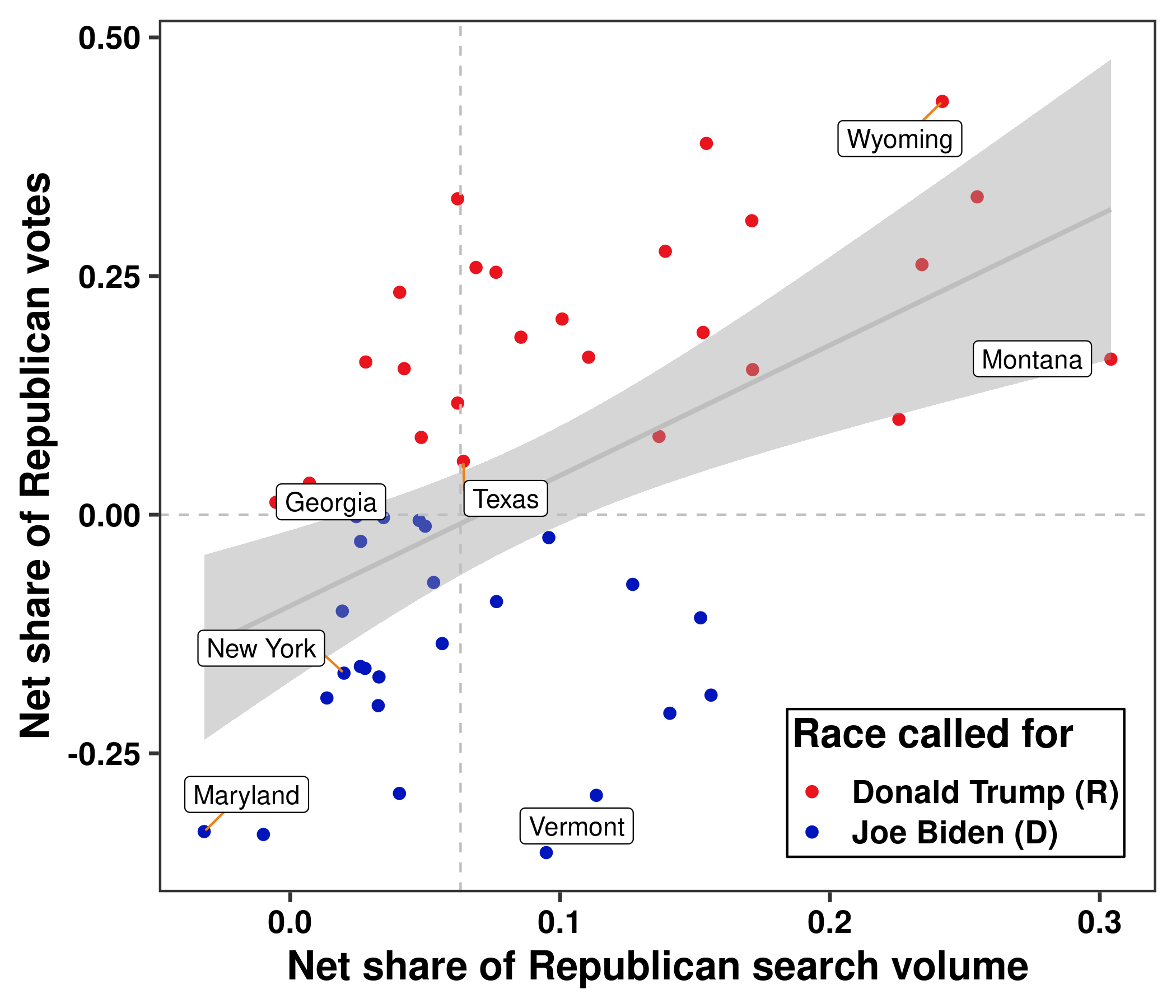}
\caption{Republican Google usage measure and Republican vote shares per state. The scatter plot displays the raw correlation between the ideological position of a state based on its population's use of partisan search terms (net Republican search volume) and the corresponding state's net share of Republican votes (share of R votes - share of D votes) in the 2020 US presidential elections. Dots of states where the US presidential race was called for Joe Biden are blue, those where the race was called for Donald Trump are red. The grey line indicates the intercept and slope coefficient from regressing the net share of Republican voters on the net Republican search volume. The gray area indicates the corresponding confidence band at a 95 percent confidence level. The dashed vertical line indicates the median net Republican search volume, the horizontal dashed line indicates a Republican net vote share of 0. Election results data are from the Federal Election Commission (FEC). Data on search volume are collected from Google Trends.}
\label{fig:searchterms_validation} 
\end{center}
\end{figure}

\clearpage
\begin{figure}[p]
\begin{center}
   \includegraphics[width=0.85\textwidth]{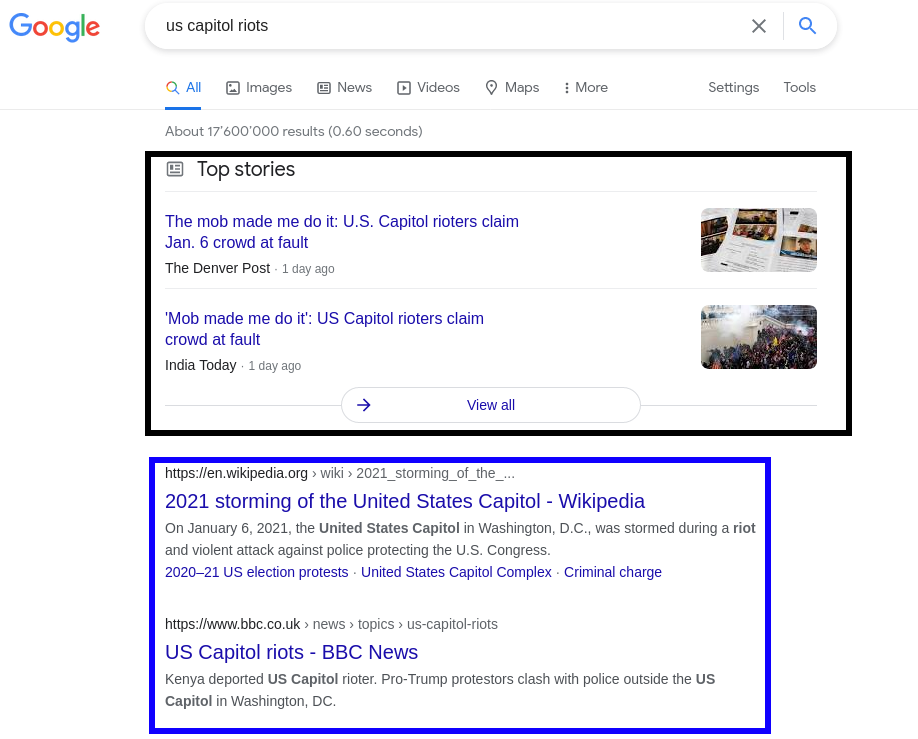}
\caption{Google search results components. Screenshot of a typical Google search results page as they occur in our data. For all election-related search results, we extract and parse the list of organic search results (blue rectangle) as well as the top stories (black rectangle).}
\label{fig:google_serp} 
\end{center}
\end{figure}

\clearpage
\begin{figure}[p]
\begin{center}
\includegraphics[width=0.49\textwidth]{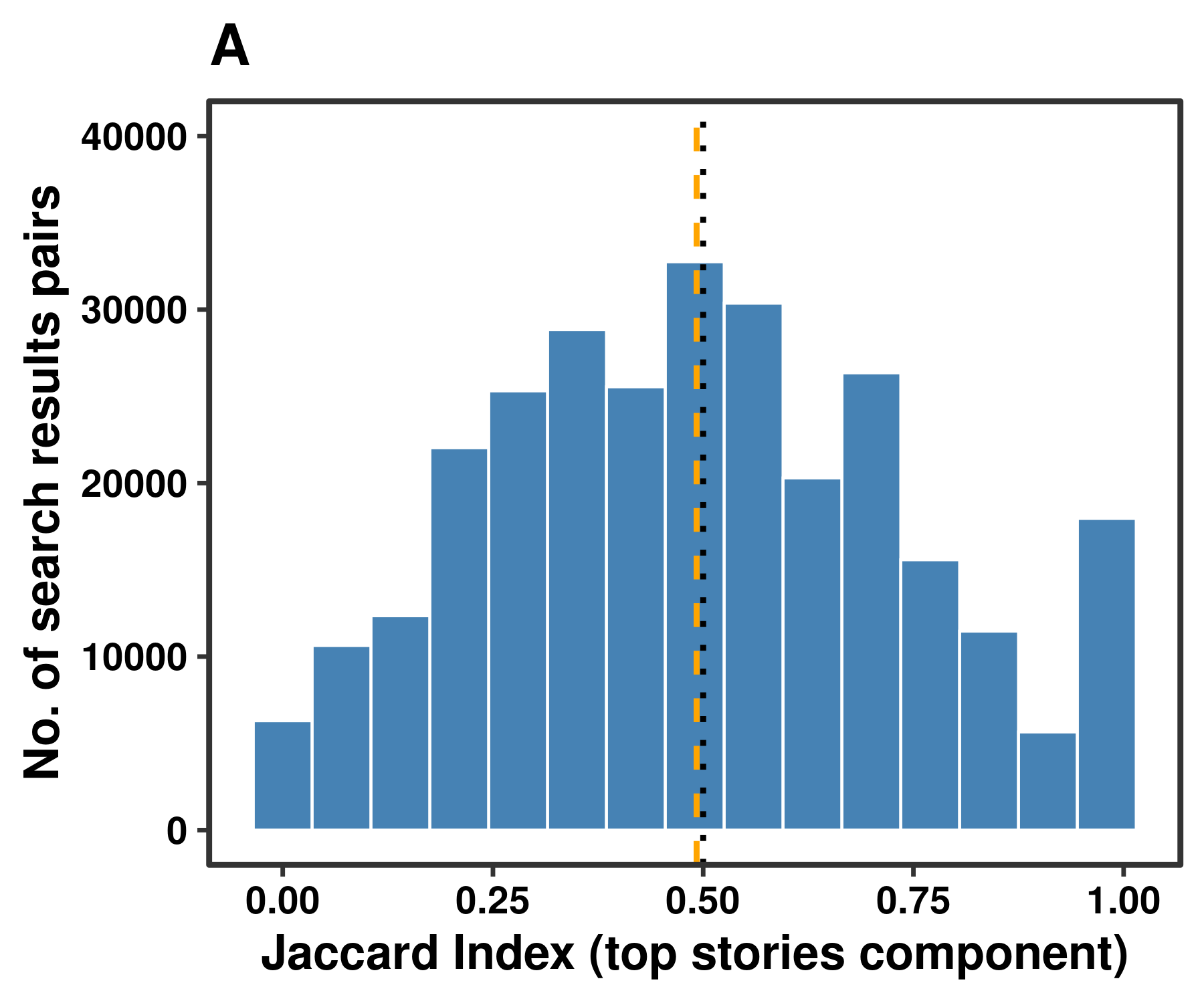}
\includegraphics[width=0.49\textwidth]{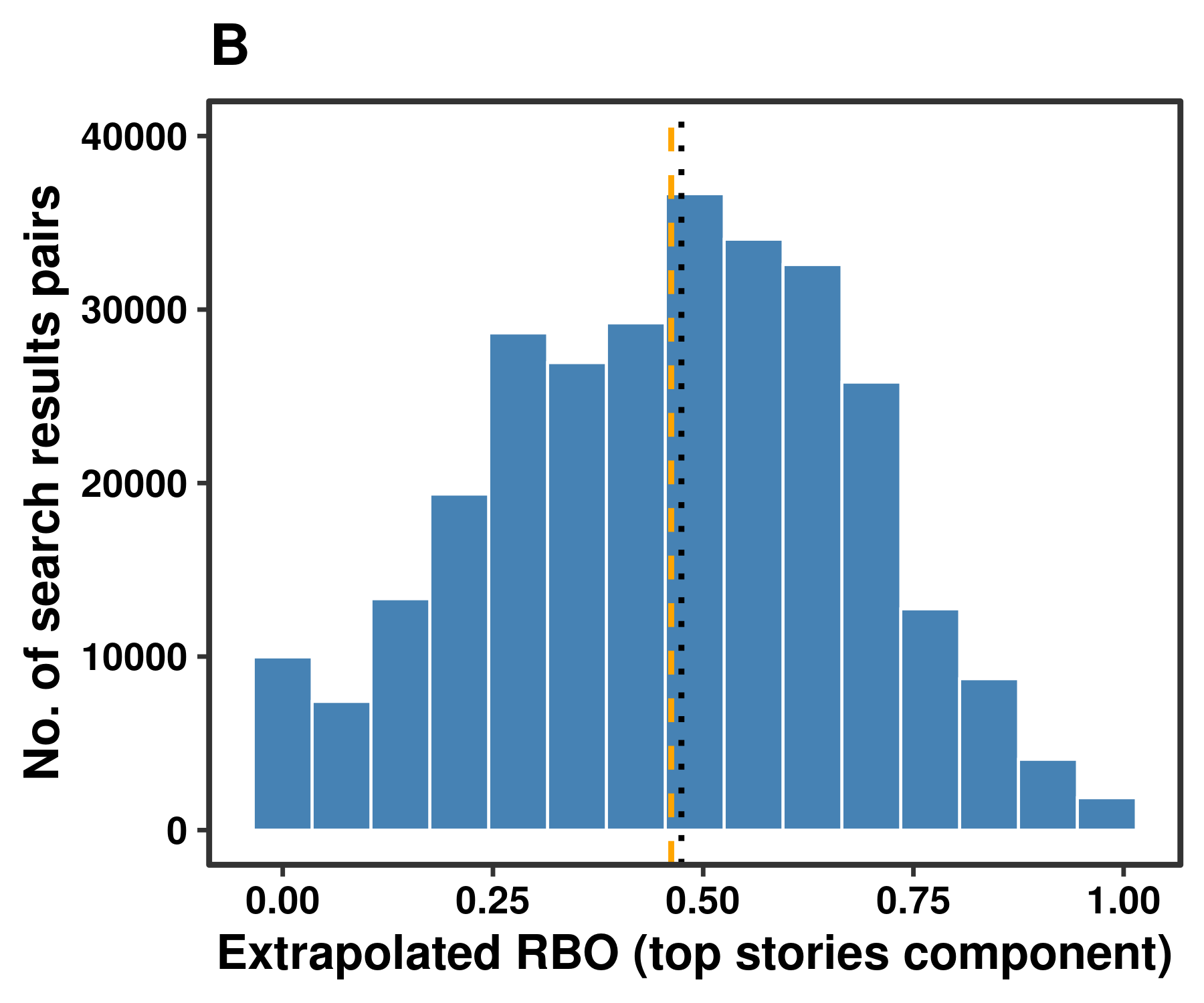}
\end{center}
\caption{Similarity of search result top stories component across user pairs. The histograms display the distributions of search results similarity in top stories components for each synthetic user-pair resulting from the same election-related queries (for which a top stories component was shown) on the same day, measured with the Jaccard Index in panel A and the extrapolated RBO in panel B. 
The dotted black (dashed orange) vertical lines indicate the median (mean) of the corresponding distribution.} 
\label{fig:serp_similarity_topstories}
\end{figure}

\clearpage
\begin{figure}
\begin{center}
\includegraphics[width=0.98\textwidth]{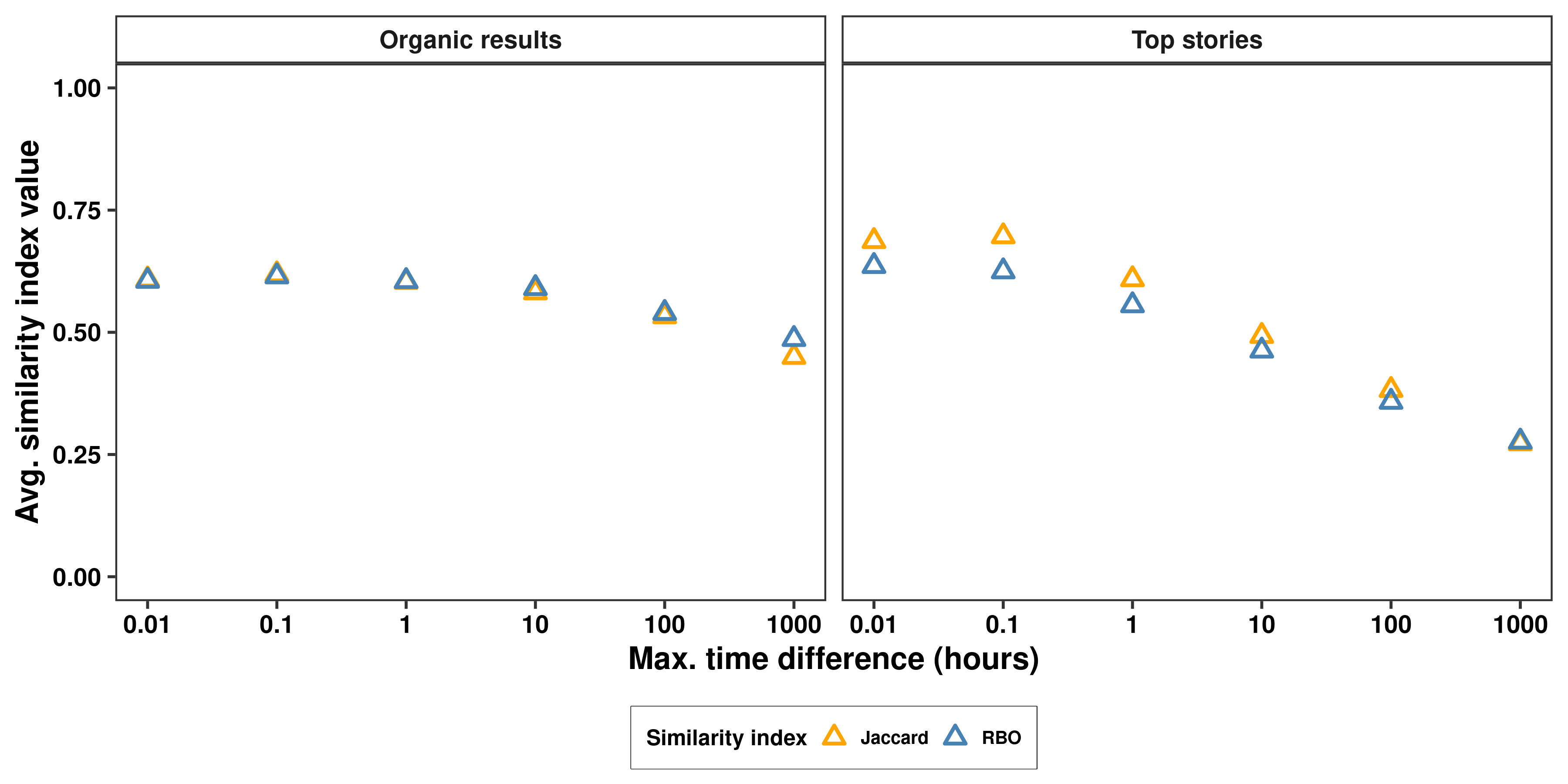}
\caption{Search results similarity and the time difference between searches. The plots display the average search results similarity (Jaccard and extrapolated RBO indices) for different search results-pair sub-samples for both organic search results (left panel) and top stories (right panel). Sub-samples are generated based on a varying maximum duration threshold (in hours) between the queries of user $i$ and user $j$ using the same election-related search term. The thresholds to create sub-samples are indicated on the horizontal axis, the corresponding average index values are indicated on the vertical axis. Reading example: the average extrapolated RBO similarity between search results resulting from a Google search based on the same search term that were issued within 0.01 hours or less is 0.61 for organic search results and 0.64 for the top stories component.}
\label{fig:timediff}
\end{center}
\end{figure}

\clearpage
\begin{figure}[p]
\begin{center}
\includegraphics[width=0.99\textwidth]{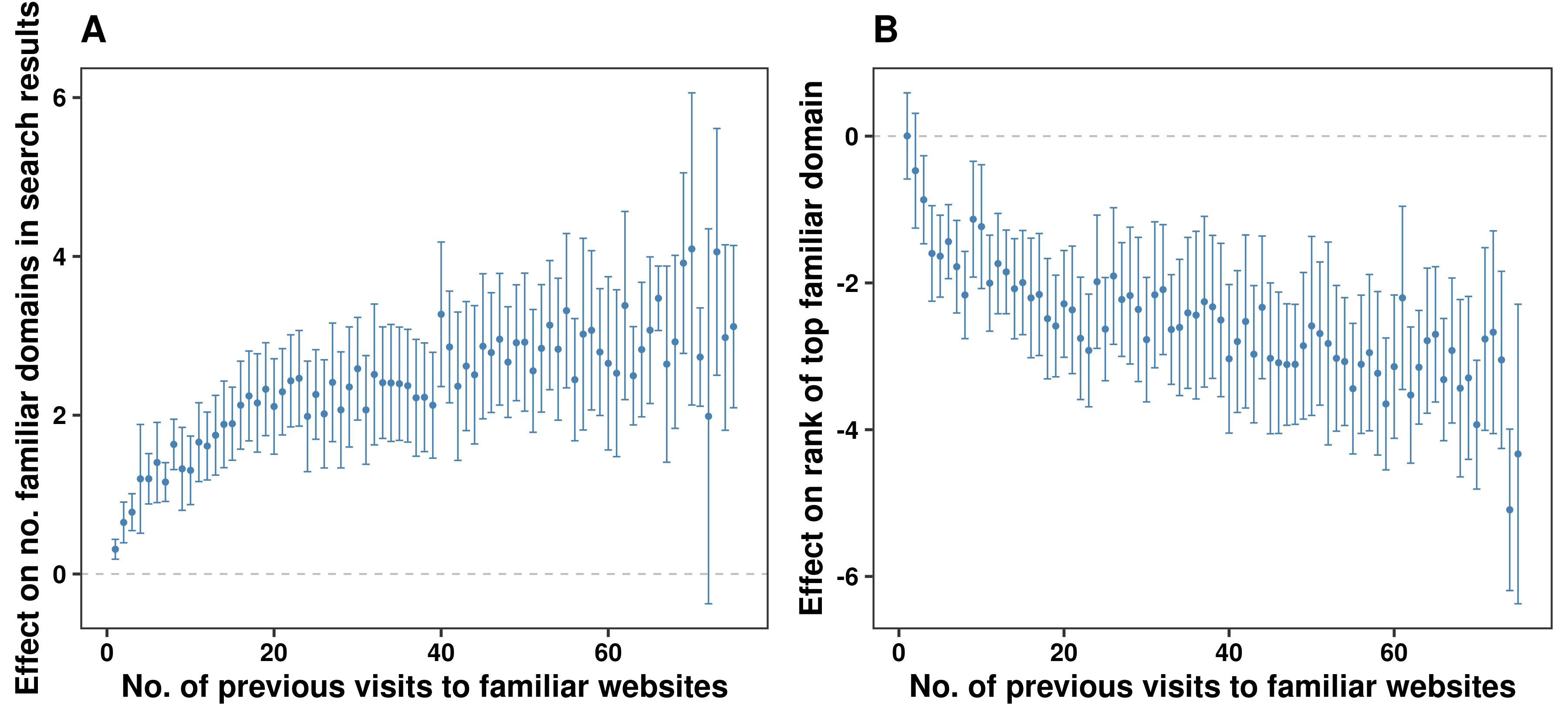}
\end{center}
\caption{Effects of the number of previous visits to familiar websites on the familiarity of the first search results page's top stories component. The dependent variable is the number of familiar domains in the search results' top story component in panel A, and the rank of the search results' top story component's top familiar domain in panel B (where we restrict the sample to search results that contain at least one of the corresponding users' familiar domains). Blue dots display marginal effects estimated from regressing the dependent variables on a set of indicator variables, one for each value of \emph{No.\ of previous visits to familiar websites} (with 0 visits as reference category), accounting for date-of-search, search-term, and browser-language fixed effects.  
Blue bars indicate 95\% confidence intervals based on standard errors three-way clustered by date of search, search term, and user.  
Extreme outliers (with values in the top 0.5\% of the number of previous visits) are excluded.}
\label{fig:familiarity_topstories}
\end{figure}

\clearpage
\begin{figure}[p]
\begin{center}
\includegraphics[width=0.99\textwidth]{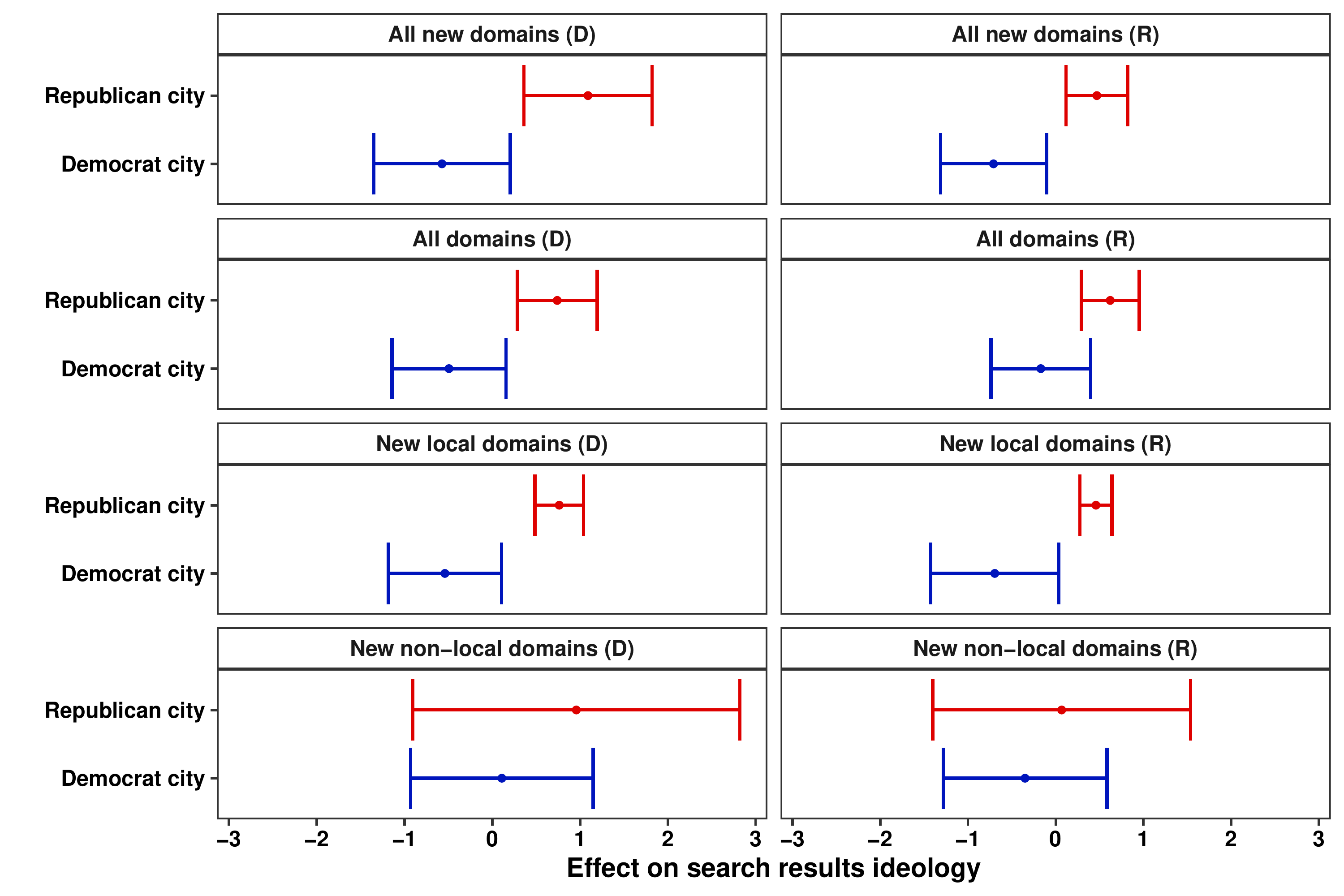}
\caption{Effects of location on search results ideology of new domains for Democrat users (left panels) and Republican users (right panels) for different sets of domains (in analogy to Figure~4 in the main text). 
Dots indicate marginal effects estimated from regressing the Search Results Ideology Score (SRIS) on indicator variables for the partisanship of the city where the user is located (with `purple' cities as reference categories), accounting for date-of-search, search-term, and browser-language fixed effects. The indicated 95\% confidence intervals are based on standard errors three-way clustered by date of search, search term, and user.}
\label{fig:city_samplesplit}
\end{center}
\end{figure}

\clearpage
\section{Tables}

\begin{table}[h]
\centering


\begin{tabular}{ccc}

  \toprule
City (user location) & GOP vote share (in \%) & City ideology \\ 
  \midrule
CA-SAN FRANCISCO & 13.40 & D \\ 
  WI-MADISON & 20.00 & D \\ 
  CA-SAN JOSE & 20.90 & D \\ 
  DC-WASHINGTON & 23.10 & D \\ 
  TX-EL PASO & 26.50 & D \\ 
  WA-SEATTLE & 28.50 & D \\ 
  IL-CHICAGO & 29.60 & D \\ 
  OR-PORTLAND & 30.40 & D \\ 
  CA-STOCKTON & 31.40 & D \\ 
  MA-BOSTON & 32.20 & D \\ 
  NY-NEW YORK & 33.30 & D \\ 
  CA-SAN DIEGO & 37.90 & purple \\ 
  CO-DENVER & 39.00 & purple \\ 
  MI-DETROIT & 41.70 & purple \\ 
  TX-HOUSTON & 44.20 & purple \\ 
  PA-HARRISBURG & 49.80 & purple \\ 
  AZ-PHOENIX & 50.70 & purple \\ 
  WA-SPOKANE & 51.10 & purple \\ 
  OH-DAYTON & 53.60 & R \\ 
  NE-OMAHA & 53.90 & R \\ 
  FL-JACKSONVILLE & 54.20 & R \\ 
  CA-BAKERSFIELD & 55.00 & R \\ 
  KS-WICHITA & 57.20 & R \\ 
  OK-OKLAHOMA CITY & 58.90 & R \\ 
  CO-COLORADO SPRINGS & 59.60 & R \\ 
    \bottomrule
    
\end{tabular}

\caption{Synthetic user locations and city ideology. List of all US cities where synthetic users were located via residential proxies. The middle column shows the city-level Republican vote shares (in percent) in the 2016 elections, and the right column the corresponding categorization into Democrat, Republican, and `purple` cities. \emph{Data:} City-level GOP vote shares from FiveThirtyEight (\url{https://projects.fivethirtyeight.com/republicans-democrats-cities/}).} 
\label{tab:sm_cities}
\end{table}
\vspace*{\fill}

\begin{table}[p]
\centering


\begin{tabular}{cc}
\toprule
Predominantly Democrat & Predominantly Republican \\ 
  \midrule
actblue.com & 48.pm \\
boingboing.net & bitchute.com \\
cnn.com & bizpacreview.com \\
crooksandliars.com & cmun.it \\
dailydot.com & conservativefiringline.com \\
dailykos.com & cosiskey.com \\
huffingtonpost.com & dailycaller.com \\
msn.com & hotpagenews.com \\
nbcnews.com & jenke.rs \\
nytimes.com & magapill.com \\
politico.com & oddcrimes.com \\
politicususa.com & pscp.tv \\
rawstory.com & robinspost.com \\
slate.com & storiesflow.com \\
splinternews.com & tacticalinvestor.com  \\
talkingpointsmemo.com & titrespresse.com \\
thehill.com & trump-news.today \\
thinkprogress.org & ussanews.com \\
vox.com & vipscandals.com \\
washingtonpost.com & wnd.com \\

   \bottomrule
\end{tabular}

\caption{Top-20 most partisan domains predominantly used by Democrats and Republicans (alphabetically ordered).} 
\label{tab:partisandomains}
\end{table}

\begin{table}[p]
\centering

\begin{tabular}{cc}
  \toprule
Predominantly Democrat & Predominantly Republican \\ 
  \midrule
affordable health & aborted babies \\ 
  air pollution & al qaeda \\ 
  attorney general & american taxpayers \\ 
  campaign finance & arms embargo \\ 
  coast guard & big government \\ 
  equal justice & bill rights \\ 
  equal rights & congressional budget \\ 
  funding department & country illegally \\ 
  hiv aids & death tax \\ 
  institutes health & discretionary spending \\ 
  investments education & farm bureau \\ 
  job training & federal bureaucrats \\ 
  labor environmental & federal debt \\ 
  lgbt americans & federal mandates \\ 
  medical research & federal spending \\ 
  oil companies & foreign policy \\ 
  oil spill & funding military \\ 
  pay equal & government spending \\ 
  people color & hardearned tax \\ 
  poverty line & health savings \\ 
  preventive health & interest debt \\ 
  puerto rican & local control \\ 
  reduce deficit & national guard \\ 
  right choose & patent rights \\ 
  shut government & private property \\ 
  tobacco companies & raise taxes \\ 
  tobacco industry & rules regulations \\ 
  vote floor & states army \\ 
  voter registration & state department \\ 
  workers rights & taxpayer funds \\ 
   \bottomrule
\end{tabular}

\caption{Top-30 most partisan bigrams predominantly used by Democrats and Republicans (alphabetically ordered).}
\label{tab:partisanbigrams}
\end{table}

\begin{landscape}
\begin{table}[!htb] \centering 

\scriptsize 

\begingroup 
\scriptsize 
\begin{tabular}{@{\extracolsep{5pt}}lcccccccccc} 
\\[-1.8ex]
\toprule \\
 & \multicolumn{10}{c}{Dependent variable:} \\ 
\cline{2-11} 
\\[-1.8ex] & \multicolumn{8}{c}{$\#$ familiar domains} & P(any familiar) & Top rank \\ 
\\[-1.8ex] & (1) & (2) & (3) & (4) & (5) & (6) & (7) & (8) & (9) & (10)\\ 
\midrule \\
 $\# Visits$ & 0.008$^{***}$ & 0.019$^{***}$ & 0.023$^{***}$ & 0.023$^{***}$ & 0.022$^{***}$ &  & 0.026$^{***}$ & 0.019$^{***}$ & 0.003$^{***}$ & $-$0.014$^{***}$ \\ 
  & (0.001) & (0.004) & (0.004) & (0.004) & (0.004) &  & (0.006) & (0.003) & (0.001) & (0.003) \\ 
  & & & & & & & & & & \\ 
 Days elapsed &  &  &  &  & $-$0.013$^{**}$ &  &  &  &  &  \\ 
  &  &  &  &  & (0.005) &  &  &  &  &  \\ 
  & & & & & & & & & & \\ 
 $\# Searches$ &  &  &  &  &  & 0.020$^{***}$ & $-$0.004 &  &  &  \\ 
  &  &  &  &  &  & (0.005) & (0.006) &  &  &  \\ 
  & & & & & & & & & & \\ 
 Constant & 1.752$^{***}$ &  &  &  &  &  &  &  &  &  \\ 
  & (0.022) &  &  &  &  &  &  &  &  &  \\ 
  & & & & & & & & & & \\ 
\midrule \\
Language FE &  & X & X & X & X & X & X & X & X & X \\ 
Search term FE &  & X &  & X & X & X & X & X & X & X \\ 
Date of search FE &  &  & X & X &  & X & X & X & X & X \\ 
Clust. SE user  &  & X & X & X & X & X & X & X & X & X \\ 
Clust. SE search term &  & X &  & X & X & X & X & X & X & X \\ 
Clust. SE date  &  &  & X & X &  & X & X & X & X & X \\ 
Observations & 24,853 & 24,853 & 24,853 & 24,853 & 24,853 & 24,853 & 24,853 & 21,379 & 24,853 & 21,379 \\ 
R$^{2}$ & 0.009 & 0.247 & 0.190 & 0.278 & 0.253 & 0.266 & 0.278 & 0.240 & 0.123 & 0.430 \\ 
Adjusted R$^{2}$ & 0.008 & 0.245 & 0.186 & 0.272 & 0.251 & 0.261 & 0.273 & 0.234 & 0.117 & 0.425 \\ 
\bottomrule\\

\end{tabular} 
\endgroup 

  \caption{Familiarity of organic search results and previous website visits. OLS regressions with heteroskedasticity robust (clustered) standard errors. Only election-related search results are considered, and extreme outliers (with values in the top 0.5\% of the number of previous visits) are excluded (consistent with Figure~3 of the research article). $\# Visits$ measures the total number of times user $i$ has previously visited familiar websites whose domains occur at least once in any election-related search results. $\# Searches$ measures the total number of times user $i$ has previously used familiar domains that ever occur on any search results as a search term. \emph{Days elapsed} measures the number of days passed since the beginning of our study (mid-October 2020) at the time the search query was issued. The dependent variable is the number of familiar domains in the search results page to election-related queries in columns (1)-(8), an indicator equal to 1 if the search results page contains any domain the user is familiar with (and 0 otherwise) in column (9), and the rank of the highest ranking domain familiar to the user in the search results (smaller values indicate higher placement in the search results) in column (10). In columns (8) and (10), the sample is restricted to observations where users see at least one familiar domain. Fixed effects and standard error clustering levels are indicated for each specification. The number of clusters are: 150 unique users, 68 unique search terms, and 113 dates for the specifications based on the full sample, and 150 unique users, 58 unique search terms, and 107 dates for for the restricted sample (columns (8) and (10)). The statistical signifiMethodscance of coefficient estimates is indicated as follows: $^{*}$p$<$0.1; $^{**}$p$<$0.05; $^{***}$p$<$0.01.} 
  \label{tab:serp_familiarity} 
\end{table} 
\end{landscape}

\begin{landscape}
\begin{table}[!htb] \centering 
 
\scriptsize 

\begingroup 
\scriptsize 
\begin{tabular}{@{\extracolsep{5pt}}lcccccccccc} 
\\[-1.8ex]
\toprule \\
 & \multicolumn{10}{c}{Dependent variable:} \\ 
\cline{2-11} 
\\[-1.8ex] & \multicolumn{8}{c}{$\#$ familiar domains} & P(any familiar) & Top rank \\ 
\\[-1.8ex] & (1) & (2) & (3) & (4) & (5) & (6) & (7) & (8) & (9) & (10)\\ 
\midrule \\[-1.8ex] 
 $\# Visits$ & 0.028$^{***}$ & 0.028$^{***}$ & 0.029$^{***}$ & 0.028$^{***}$ & 0.028$^{***}$ &  & 0.053$^{***}$ & 0.026$^{***}$ & 0.003$^{***}$ & $-$0.025$^{***}$ \\ 
  & (0.001) & (0.005) & (0.005) & (0.006) & (0.006) &  & (0.009) & (0.005) & (0.001) & (0.005) \\ 
  & & & & & & & & & & \\ 
 Days elapsed &  &  &  &  & $-$0.0002 &  &  &  &  &  \\ 
  &  &  &  &  & (0.005) &  &  &  &  &  \\ 
  & & & & & & & & & & \\ 
 $\# Searches$ &  &  &  &  &  & 0.020$^{***}$ & $-$0.029$^{***}$ &  &  &  \\ 
  &  &  &  &  &  & (0.007) & (0.010) &  &  &  \\ 
  & & & & & & & & & & \\ 
 Constant & 1.604$^{***}$ &  &  &  &  &  &  &  &  &  \\ 
  & (0.035) &  &  &  &  &  &  &  &  &  \\ 
  & & & & & & & & & & \\ 
\midrule \\ 
Language FE &  & X & X & X & X & X & X & X & X & X \\ 
Search term FE &  & X &  & X & X & X & X & X & X & X \\ 
Date of search FE &  &  & X & X &  & X & X & X & X & X \\ 
Clust. SE user  &  & X & X & X & X & X & X & X & X & X \\ 
Clust. SE search term &  & X &  & X & X & X & X & X & X & X \\ 
Clust. SE date  &  &  & X & X &  & X & X & X & X & X \\ 
Observations & 8,626 & 8,626 & 8,626 & 8,626 & 8,626 & 8,626 & 8,626 & 6,726 & 8,626 & 6,726 \\ 
R$^{2}$ & 0.053 & 0.279 & 0.205 & 0.314 & 0.279 & 0.292 & 0.322 & 0.310 & 0.413 & 0.209 \\ 
Adjusted R$^{2}$ & 0.053 & 0.273 & 0.194 & 0.300 & 0.273 & 0.278 & 0.308 & 0.292 & 0.401 & 0.189 \\ 
\bottomrule \\
\end{tabular} 
\endgroup 

 \caption{Familiarity of search result top stories components and previous website visits. OLS regressions with heteroskedasticity robust (clustered) standard errors. The unit of observation is user-search results page. Only the top stories components to election-related queries are considered, and extreme outliers (with values in the top 0.5\% of the number of previous visits) are excluded (consistent with Figure~\ref{fig:familiarity_topstories}). $\# Visits$ measures the total number of times user $i$ has previously visited familiar websites whose domains occur at least once in any election-related search results. $\# Searches$ measures the total number of times user $i$ has previously used familiar domains that ever occur on any search results as a search term. \emph{Days elapsed} measures the number of days passed since the beginning of our study (mid-October 2020) at the time the search query was issued. The dependent variable is the number of familiar domains in the search results page to election-related queries in columns (1)-(8), an indicator equal to 1 if the search results page contains any domain the user is familiar with (and 0 otherwise) in column (9), and the rank of the highest ranking domain familiar to the user in the search results (smaller values indicate higher placement in the search results) in column (10). In columns (8) and (10), the sample is restricted to observations where users see at least one familiar domain. Fixed effects and standard error clustering levels are indicated for each specification. The number of clusters are: 150 unique users, 68 unique search terms, and 113 dates for the specifications based on the full sample, and 150 unique users, 58 unique search terms, and 107 dates for for the restricted sample (columns (8) and (10)). The statistical significance of coefficient estimates is indicated as follows: $^{*}$p$<$0.1; $^{**}$p$<$0.05; $^{***}$p$<$0.01.} 
  \label{tab:serp_familiarity_topstories} 
\end{table} 

\end{landscape}

\begin{landscape}
\begin{table}[!htb] \centering 

\scriptsize

\begingroup 
\small 
\begin{tabular}{@{\hspace{5pt}}l@{\hspace{5pt}}ccccccccc} 
\toprule 
 & \multicolumn{9}{c}{Dependent variable: search results ideology score} \\ 
\cmidrule(rr){2-10} 
 & (1) & (2) & (3) & (4) & (5) & (6) & (7) & (8) & (9)\\ 
\midrule  
\\[-2.1ex] Democrat & $-$0.162 & $-$0.229$^{**}$ & $-$0.134 & $-$0.103 & $-$0.210$^{***}$ & $-$0.141 & 0.108 & 0.128$^{*}$ & $-$0.154 \\ 
  & (0.129) & (0.095) & (0.105) & (0.189) & (0.058) & (0.096) & (0.102) & (0.074) & (0.107) \\ 
 \addlinespace 
 Republican & $-$0.175 & $-$0.190$^{*}$ & $-$0.166 & 0.151 & $-$0.159 & $-$0.091$^{***}$ & 0.177$^{**}$ & 0.119 & $-$0.092 \\ 
  & (0.131) & (0.109) & (0.117) & (0.179) & (0.134) & (0.030) & (0.088) & (0.097) & (0.083) \\ 
 \addlinespace 
 Dem. city & $-$0.554 & $-$0.477 & $-$0.534 & 0.021 & $-$0.227 & $-$0.242$^{**}$ & $-$0.464$^{***}$ & $-$0.580$^{**}$ & $-$0.515 \\ 
  & (0.333) & (0.346) & (0.354) & (0.243) & (0.155) & (0.113) & (0.170) & (0.273) & (0.391) \\ 
 \addlinespace 
 Rep. city & 0.633$^{***}$ & 0.527$^{***}$ & 0.501$^{***}$ & 0.492 & 0.199$^{**}$ & 0.146$^{***}$ & 0.049 & $-$0.151$^{***}$ & 0.728$^{***}$ \\ 
  & (0.185) & (0.184) & (0.124) & (0.460) & (0.095) & (0.045) & (0.052) & (0.048) & (0.263) \\ 
 \addlinespace 
\midrule  
Set of domains & All new & All & New local & New non-local & All new & All new & All new & All new & All new \\ 
Ideology indices & All 5 & All 5 & All 5 & All 5 & Bakshy et al. & Budak & Mturk & Pew & Robertson et al. \\ 
\% domains in index & 79.15 & 79.15 & 79.15 & 79.15 & 36.62 & 7.2 & 43.26 & 15.16 & 77.96 \\ 
Language FE & X & X & X & X & X & X & X & X & X \\ 
Search term FE & X & X & X & X & X & X & X & X & X \\ 
Date of search FE & X & X & X & X & X & X & X & X & X \\ 
Observations & 25,644 & 25,644 & 25,543 & 11,755 & 25,644 & 25,644 & 25,644 & 25,644 & 25,644 \\ 
R$^{2}$ & 0.776 & 0.784 & 0.799 & 0.839 & 0.613 & 0.685 & 0.657 & 0.616 & 0.775 \\ 
Adjusted R$^{2}$ & 0.774 & 0.783 & 0.798 & 0.837 & 0.610 & 0.682 & 0.654 & 0.613 & 0.774 \\ 
\bottomrule 

\end{tabular} 
\endgroup 

  \caption{Effects of partisanship and location on search results ideology of new domains in the organic search results. OLS regressions with robust standard errors three-way clustered by synthetic user, search term, and date of search. The unit of observation is user-search results page. The dependent variable is the Search Result Ideology Score (SRIS, see equation 3 in the research article) for different sets of domains listed in the organic search results to election-related queries: All new domains (i.e., domains not initially known by any of the synthetic users) in columns (1) and (5)-(9), all (new and known) domains in column (2), new local domains in column (3), and new non-local domains in column (4). SRIS is based on the five different website ideology indices in columns (1)-(4) and on a single website ideology index indicated under ``Ideology index'' in columns (5)-(9). In all cases, SRIS is measured on a liberal-conservative scale [-100,100]. \emph{Democrat} (\emph{Republican}) is equal to 1 if a synthetic user is configured as having Democrat (Republican) browsing and search preferences. The reference category is a non-partisan synthetic user (no partisan browsing and search behavior). \emph{Dem. city} (\emph{Rep. city}) is equal to 1 if a synthetic user is located in a city with predominantly Democratic (Republican) voters. The reference category is purple cities. In all columns we account for browser language fixed effects, search term fixed effects, and date of search fixed effects. The statistical significance of coefficient estimates is indicated as follows: $^{*}$p$<$0.1; $^{**}$p$<$0.05; $^{***}$p$<$0.01.} 
  \label{tab:serp_ideology_notknown} 
\end{table} 
\end{landscape}
\clearpage

\begin{table}[p]
\centering

\begin{tabular}{llrr}
  \toprule
Specification & Hypothesis & Chisq & Pr($>$Chisq) \\ 
  \midrule
(1) & $\hat{\beta}_{Democrat}-\hat{\beta}_{Repuclican}=0$ & 0.01 & 0.93 \\ 
  (2) & $\hat{\beta}_{Democrat}-\hat{\beta}_{Repuclican}=0$ & 0.13 & 0.72 \\ 
  (3) & $\hat{\beta}_{Democrat}-\hat{\beta}_{Repuclican}=0$ & 0.07 & 0.79 \\ 
  (4) & $\hat{\beta}_{Democrat}-\hat{\beta}_{Repuclican}=0$ & 0.99 & 0.32 \\ 
  (5) & $\hat{\beta}_{Democrat}-\hat{\beta}_{Repuclican}=0$ & 0.13 & 0.71 \\ 
  (6) & $\hat{\beta}_{Democrat}-\hat{\beta}_{Repuclican}=0$ & 0.17 & 0.68 \\ 
  (7) & $\hat{\beta}_{Democrat}-\hat{\beta}_{Repuclican}=0$ & 0.47 & 0.49 \\ 
  (8) & $\hat{\beta}_{Democrat}-\hat{\beta}_{Repuclican}=0$ & 0.01 & 0.94 \\ 
  (9) & $\hat{\beta}_{Democrat}-\hat{\beta}_{Repuclican}=0$ & 0.51 & 0.48 \\ 
  (1) & $\hat{\beta}_{Dem. city}-\hat{\beta}_{Rep. city}=0$ & 6.41 & 0.01 \\ 
  (2) & $\hat{\beta}_{Dem. city}-\hat{\beta}_{Rep. city}=0$ & 4.17 & 0.04 \\ 
  (3) & $\hat{\beta}_{Dem. city}-\hat{\beta}_{Rep. city}=0$ & 5.37 & 0.02 \\ 
  (4) & $\hat{\beta}_{Dem. city}-\hat{\beta}_{Rep. city}=0$ & 0.80 & 0.37 \\ 
  (5) & $\hat{\beta}_{Dem. city}-\hat{\beta}_{Rep. city}=0$ & 9.98 & 0.00 \\ 
  (6) & $\hat{\beta}_{Dem. city}-\hat{\beta}_{Rep. city}=0$ & 9.78 & 0.00 \\ 
  (7) & $\hat{\beta}_{Dem. city}-\hat{\beta}_{Rep. city}=0$ & 15.70 & 0.00 \\ 
  (8) & $\hat{\beta}_{Dem. city}-\hat{\beta}_{Rep. city}=0$ & 1.82 & 0.18 \\ 
  (9) & $\hat{\beta}_{Dem. city}-\hat{\beta}_{Rep. city}=0$ & 4.13 & 0.04 \\ 
   \bottomrule
\end{tabular}

\caption{Linear hypothesis tests based on the estimates presented in Table~\ref{tab:serp_ideology_notknown}. Column `Hypothesis' indicates the form of the linear hypothesis tests. The tests aim to clarify whether the estimated beta coefficients of a Democrat user (Democrat city) indicator is statistically significantly different from the beta coefficients of a Republican user (Republican city). The null in all tests is that the two coefficients are of the same size. The corresponding p-values are shown in the right-most column.} 
\label{tab:serp_ideology_linh_notknown}
\end{table}

\begin{table}[p]
\centering

\begin{tabular}{@{\hspace{5pt}}l@{\hspace{5pt}}cccc} 
\toprule 
 & \multicolumn{4}{c}{Dependent variable: search results ideology score} \\ 
\cmidrule(rr){2-5} 
 & (1) & (2) & (3) & (4)\\ 
\midrule  
\\[-2.1ex] Visited sites ideology & 0.002 & 0.0001 & 0.001 & 0.023 \\ 
  & (0.007) & (0.006) & (0.005) & (0.017) \\ 
 \addlinespace 
 Share Rep. voters in city & 0.030$^{**}$ & 0.026 & 0.026$^{*}$ & 0.013 \\ 
  & (0.015) & (0.015) & (0.014) & (0.015) \\ 
 \addlinespace 
\midrule  
Set of domains & All new & All & New local & New non-local \\ 
Observations & 25,644 & 25,644 & 25,543 & 11,755 \\ 
R$^{2}$ & 0.775 & 0.784 & 0.799 & 0.839 \\ 
Adjusted R$^{2}$ & 0.774 & 0.783 & 0.797 & 0.837 \\ 
\bottomrule 

\end{tabular} 

\caption{Effects of continuous measures of user and city ideology on search results ideology. OLS regressions with robust standard errors three-way clustered by synthetic user, search term, and date of search. The dependent variable is the Search Result Ideology Score (SRIS, see equation 3 in the research article) for different sets of domains listed in the organic search results to election-related queries: All new domains (i.e., domains not initially known by any of the synthetic users) in column (1), all (new and known) domains in column (2), new local domains in column (3), and new non-local domains in column (4).
\emph{Visited Sites Ideology} is the average ideology score of web domains visited by the synthetic user.  \emph{Share Rep.\ voters} is the share of Republican voters (in \%) in the city where the synthetic user is located. The differences in the number of observations in columns (3) and (4) is due to the fact that some search results pages only contain non-local websites and some only local websites, leading to missing values in the dependent variable in some cases. In all columns we account for browser language fixed effects, search term fixed effects, and date of search fixed effects. The statistical significance of coefficient estimates is indicated as follows: $^{*}$p$<$0.1; $^{**}$p$<$0.05; $^{***}$p$<$0.01.} \label{tab:serp_ideology_userideology}
\end{table}




\FloatBarrier





\bibliographystyle{apalike}
\bibliography{sn-bibliography}